\documentclass[acmsmall]{acmart} 

\AtBeginDocument{%
  \providecommand\BibTeX{{%
    \normalfont B\kern-0.5em{\scshape i\kern-0.25em b}\kern-0.8em\TeX}}}

\setcopyright{acmcopyright}
\acmJournal{PACMHCI}
\acmYear{2022} \acmVolume{6} \acmNumber{CSCW2} \acmArticle{496} \acmMonth{11} \acmPrice{15.00}\acmDOI{10.1145/3555597}

\usepackage{color}
\usepackage{booktabs}
\usepackage{csquotes}
\usepackage{multirow}
\usepackage{makecell}

\newcommand{\qian}[1]{{\color{black} #1}}
\newcommand{\minor}[1]{{\color{black} #1}}

\newcommand{\eone}{\includegraphics[scale=0.06]{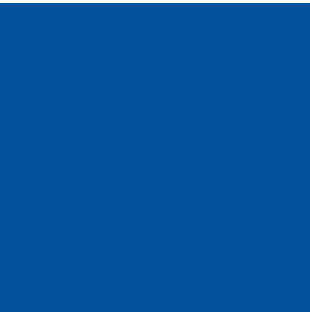}}
\newcommand{\efive}{\includegraphics[scale=0.06]{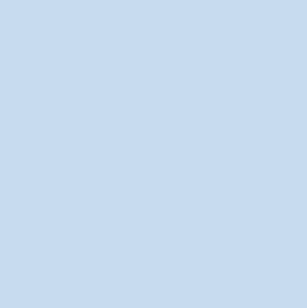}}

\begin{document}

\title{\qian{Bias-Aware} Design for Informed Decisions: Raising Awareness of Self-Selection Bias in User Ratings and Reviews}


\author{Qian Zhu}
\affiliation{%
  \institution{The Hong Kong University of Science and Technology}
  \city{Hong Kong}
  \country{China}}
\email{qian.zhu@connect.ust.hk}

\author{Leo Yu-Ho Lo}
\affiliation{%
  \institution{The Hong Kong University of Science and Technology}
  \city{Hong Kong}
  \country{China}}
\email{leoyuholo@gmail.com}

\author{Meng Xia}
\affiliation{%
  \institution{Carnegie Mellon University}
  \city{Pittsburgh}
  \country{United States}}
\email{iris.xia@connect.ust.hk}

\author{Zixin Chen}
\affiliation{%
  \institution{The Hong Kong University of Science and Technology}
  \city{Hong Kong}
  \country{China}}
\email{zchendf@connect.ust.hk}

\author{Xiaojuan Ma}
\affiliation{%
  \institution{The Hong Kong University of Science and Technology}
  \city{Hong Kong}
  \country{China}}
\email{mxj@cse.ust.hk}






\renewcommand{\shortauthors}{Qian Z., et al.}

\begin{abstract}
People often take user ratings/reviews into consideration when shopping for products or services online. However, such user-generated data contains self-selection bias that could affect people's decisions and it is hard to resolve this issue completely by algorithms. 
In this work, we propose to raise people's awareness of the self-selection bias by making three types of information concerning user ratings/reviews transparent. We distill these three pieces of information, \textit{i.e., reviewers' experience, the extremity of emotion, and reported aspect(s)}, from the definition of self-selection bias and exploration of related literature. 
We further conduct an online survey to assess people's perceptions of the usefulness of such information and identify the exact facets (e.g., negative emotion) people care about in their decision process. 
Then, we propose a visual design to make such details behind user reviews transparent and integrate the design into an experimental website for evaluation. The results of a between-subjects study demonstrate that our \qian{bias-aware} design significantly increases people’s awareness of bias and their satisfaction \minor{with} decision-making. We further offer a series of design implications for improving information transparency and awareness of bias in user-generated content.
\end{abstract}



\maketitle

\section{Introduction}
People increasingly rely on online word of mouth (WOM) to learn about the quality of products or services when making decisions on the Internet ~\cite{Peter:12Impact}. 
Recent research showed that 87\% of consumers read online reviews and 79\% trust them as much as in-person recommendations ~\cite{Murphy:20Survey}. 
However, various kinds of bias are embedded in user-generated contents that could affect viewers' decisions ~\cite{baeza:18bias, eslami2017careful}. 
Among them, the self-selection bias widely exists in user ratings and reviews ~\cite{HU:09J-shaped, Bhole:17Self-SBias}, and it is mainly caused by people's subjective participation in rating or writing reviews online ~\cite{HU:09J-shaped, Bhole:17Self-SBias}.  
The typical reason for self-selection bias is people's tendency to give feedback only when they are extremely satisfied or unsatisfied with the product or the service they had received ~\cite{HU:09J-shaped, Bhole:17Self-SBias}. As a result, online rating/review becomes less representative ~\cite{H:20Representative} with the biased distribution of user feedback and people who are not aware of this bias may consequently be misled ~\cite{bareinboim:12causal}. Hence, it is important to raise the awareness of self-selection bias and reduce its negative impact on users' decisions. 

Some businesses try to mitigate the self-selection bias by sending emails to encourage people to leave comments, such as Yelp\footnote{\url{https://www.yelp.com}} and Tripadvisor\footnote{\label{Trip}\url{https://www.tripadvisor.com/}}. Such solicited reviews could provide a more representative set of user feedback ~\cite{H:20Representative}. However, the companies themselves also generated biased reviews through the creation of promotional user responses ~\cite{May14:Promotional}. 
Another strategy adopted by companies is to ask customers to write about both the positives and negatives of their experiences or rate independently on multiple aspects of a product ~\cite{wu2017mitigating, nagtegaal2020designing}.
Although these methods can provide comprehensive information for users, they mostly rely on financial or social incentives and thus may not scale well ~\cite{H:20Representative}. Moreover, these solutions are proposed from the perspective of business management, which aims to control the quality of users' feedback and improve the company's reputation. They do not take the end-users' best interest into consideration, who are at the end making decisions based on the biased data.

From a technical perspective, some works reduce the bias in user-generated data by statistical removing or changing the underrepresented samples ~\cite{Flavio17:NIPS}, and other works try to mitigate the self-selection bias in user ratings/reviews by modeling, detecting and reporting suspected biases ~\cite{zheng:21identifying, zhang:19understanding}.
These approaches consider bias as a kind of feature in data, but thus far, they can not capture it perfectly ~\cite{ZHENG21:bias_usingDL}. In addition, these technical approaches are often opaque for end-users who cannot understand the output of algorithms, thus causing some distrust issues ~\cite{Emilee18:Transparency}.


\qian{The above two kinds of methods mainly focus on enhancing the sampling methods of user feedback or manipulating the data to reduce biases, but they can be difficult and costly to implement in practice. Furthermore, they can not eliminate the bias completely, which has been the nature of user-generated data.
In addition, these two methods  did not consider the perspectives of end-users who refer to the ratings/reviews to make their own decisions. For example, in the view of end-users, both strategies used by the enterprise and automatic algorithms at the back-end are invisible. Users might not realize the remaining  bias behind the data processed by these methods when viewing the statistical mean ratings to assess the quality of an online product or service. 
Instead of trying to reduce the bias in data from the back-end, in this work, we proposed to make the potential bias “visible” to end-users to help them make informed decisions. We aim to \textbf{raise people's awareness of the self-selection bias} embedded in user-generated data -- an approach to directly mitigating the impact of bias on people's decision-making  ~\cite{baeza:18bias}. }

To achieve this goal, we first propose to raise consumers' awareness of the self-selection bias in user ratings/reviews by making three types of information transparent, which are \textit{(1) the reviewers' experience, (2) the extremity of emotion, and (3) the reported aspects} in user reviews. We distilled these pieces of information according to the literature and the definition of self-selection bias ~\cite{Bhole:17Self-SBias, li:08self, Askalidis:17overcoming, gong2015characterizing}. 
Next, we conduct a large-scale survey (n = 206) to assess people's perceptions of these three types of information and identify the exact facets that are critical for their decision-making under the hotel booking scenario. 
Then, we design a visual display of these information aspects underneath user ratings/reviews and refine the design based on the feedback from a pilot study with two visualization experts and 12 users. 
We integrate the design into a prototype system that simulates typical online hotel booking experiences. 
After that, we explore how the \qian{bias-aware} design may affect users' awareness of the bias and their decision-making process through a between-subjects study. 
Experimental results show that the design can raise people’s awareness of the self-selection bias while also being thought of as helpful for making informed decisions. We discuss our findings and offer design implications at the end of the paper.

This work has three key contributions. (1) We identify three kinds of information related to the self-selection bias in online ratings/reviews and collect the specific information, including the reviewers' experience, emotion and aspects, that are needed to be visible to users. (2) We propose a \qian{bias-aware} design and integrate it into an experimental platform with real-world data. (3) We derive empirical insights from a between-subjects study and offer implications for future designs to improve transparency and awareness of bias in user-generated content.  

\section{Related Work}
Our work contributes to the existing literature on how to support decision making based on online word of mouth (WOM). We describe the relevant literature from three aspects: (1) large-scale review analysis for decision-making, (2) reducing bias in data for decision-making, and (3) transparency for raising awareness and informed decisions.
\subsection{User Review Analysis for Decision-Making}
It has become common for people to make decisions based on others' opinions posted online, such as buying a product or booking a hotel online. However, it takes a lot of time for people to dig through large numbers of online reviews ~\cite{Peter:12Impact}. 
Existing works in the natural language processing (NLP) domain proposed automatic algorithms to extract representative information and/or general user attitudes from reviews, such as generating summaries of reviews ~\cite{suhara:20opiniondigest, angelidis:21extractive, TSAI:20summari} and classifying emotions in reviews by sentiment analysis ~\cite{Binder:19ExplainingTS, He:17ApplicationOS,boiy:07automatic}. 
Recent studies tried to define the features of helpful reviews ~\cite{du:20advanced, Ali:17Emotion} by mining the factors related to user satisfaction ~\cite{Lee:19OpinionSeer} or analyzing the helpfulness by combining reviews and quantitative ratings jointly ~\cite{Chatterjee:20helpful}.

Based on these automated approaches, another body of work offered a visual interface or an interactive system for users to dig through online reviews in a more detailed manner ~\cite{Eivind:12Viz1, Chen:15VizSentiment, wu2010opinionseer}. 
They tend to present user reviews from multiple semantic perspectives and assist users in analyzing reviews from a higher level aspect. 
For example, Chang et al. visualized aspect-level data in hotel reviews for users to gain insights and analyze different types of customers ~\cite{CHANG:19Hilton}. ExtremeReader generates abstract visual summaries by offering a high-level structure of opinions in user reviews ~\cite{Xiaolan:20ExtremeReader}. Considering the subjective aspects of user reviews, recent work designed an interactive visualization system with multiple views for domain experts to mine people's opinions ~\cite{zhang:20teddy}. 
These works provide users more space to explore and customize insights from user reviews. However, most of these systems or interfaces are complex for ordinary users and they are often designed for experts to analyze large-scale data.

The above works mainly focus on how to efficiently extract and represent user reviews, while they do not give much consideration to the potential bias hidden behind them. 
Considering the pervasive subjectivity in online reviews ~\cite{halevy:19ubiquity, li:19subjective}, there is a strong need to reduce the impact of the self-selection bias on people's decision-making and help them make informed decisions. In this work, we leverage automatic approaches to analyze and extract the critical aspects of user ratings and reviews, and integrate the output into a \qian{bias-aware} design for raising people's awareness.  
\subsection{Understanding and Mitigating Bias in User-generated Data} \label{lr2.2}
Previous researchers have indicated that user ratings and reviews are biased, not only caused by the companies (i.e., the organization/person who can manipulate ratings or reviews), but also by people who gave feedback ~\cite{aral:14problem, H:20Representative, cicognani:16social}. 
One typical type of bias caused by people is the \textbf{self-selection bias}, which leads to more extreme positive and negative feedback in user ratings/reviews.
While there are several other types of bias caused by people or platform interventions, like anchoring bias, social influence bias, or rating bias by algorithms in the online community ~\cite{Dina:14Promotional, Chevalier:18Managers, thebault2017simulation}, in this paper, we mainly focus on the self-selection bias in user ratings and reviews ~\cite{Bhole:17Self-SBias, HU:09J-shaped, hu:09overcoming, H:20Representative, sterne2008addressing}.

To improve the representativeness of user feedback, researchers in the field of business and marketing designed different strategies to mitigate biases in data ~\cite{Kemal:20Manipulation, lim:17mitigating, wu:17mitigating, nagtegaal:20designing}. 
These strategies include (1) sending emails to a random selection of users and encouraging them to write reviews ~\cite{Askalidis:17overcoming, H:20Representative}, (2) offering a relative comprehensive framework for users to give feedback (e.g., commenting on the pros and cons of a subject separately) ~\cite{lim:17mitigating, wu:17mitigating}, and (3) selectively displaying a representative user feedback online by manipulating the orders ~\cite{eslami2017careful}.
However, these approaches are primarily designed for businesses with the aim of maintaining the reputation of a platform. They paid little attention to the end-users who refer to user ratings or reviews to make decisions. Additionally, biases may also be brought by such strategies, as their processes may include financial or social incentives ~\cite{H:20Representative}.

Some other researchers proposed technical methods to handle biases in user ratings and reviews ~\cite{zheng:21identifying, Riyaz:11EstimatingSeq, zhang:19understanding, calmon:17optimized, barrett:19adversarial, Walker:17TowardsMB}. These works aim to detect ~\cite{calmon:17optimized, Walker:17TowardsMB} or model ~\cite{zheng:21identifying, Riyaz:11EstimatingSeq, zhang:19understanding} biases in online ratings or reviews to mitigate the effects caused by biases. 
For instance, a recent work by Zheng et al. ~\cite{zheng:21identifying} identified and modeled biased user ratings based on textual reviews using deep learning models. 
Riyaz and Kriti used the \textit{Kalman filtering technique} to estimate the sequential bias in user reviews ~\cite{Riyaz:11EstimatingSeq}. 
There are also plenty of works that focus on building or designing unbiased algorithms in recommendation systems ~\cite{joachims2017unbiased, schnabel2016recommendations} or rating systems ~\cite{bishop2015amazon, hickey2015suspicious}, which tend to deal with biases in algorithms rather than data.
Although these approaches are proven to be useful to some extent, they still suffer issues of scalability, and they cannot reduce the bias completely due to the diversity of bias as well as the difficulty of designing unbiased algorithms ~\cite{ZHENG21:bias_usingDL}. 
Furthermore, they keep biases detected by the back-end algorithms that are opaque to end-users, and thus may cause issues of trust and explainability ~\cite{damak2021debiased}.

The self-selection bias in data are difficult to eliminate completely by automatic algorithms or data management strategies because it is caused by humans' self-selected behaviors. Additionally, the previous approaches typically performed operations that were invisible to the end-users and thus cannot help them to realize the potential bias and thus directly aid their decision making process. Therefore, it is critical to make people informed of the potential biased data \minor{while making decisions} with user rating and reviews.
In this work, we aim to raise people's awareness of the self-selection bias by putting the decision in the hands of users and reduce the risk of users being influenced by biased data rather than directly addressing biases.



\subsection{Transparency for Decision-Making and Raising Awareness}
Transparency can be used as a mechanism to help users make informed decisions with systems, and it usually comes with two goals: showing how and what ~\cite{Emilee18:Transparency, zhang:20explainable}.
For the \textbf{``how''} situation, transparency can be used to make the process of how data is processed and analyzed visible to users. 
For example, in some recommendation systems, it is easier for users to make sense of the underlying algorithms when the process recommendation is shown ~\cite{zanker:10knowledgeable, gedikli:14should}. Transparency can greatly improve users' perception of the credibility and usefulness of a system ~\cite{wang:07recommendation}, and empower the decision-making processes ~\cite{Motahhare:19Opacity, diakopoulos:16accountability}. 
The \textbf{``what''} type of transparency deals with the outcome of a system, revealing hidden data (if any) or displaying the reasons for the output ~\cite{zhang:20explainable}. 
Researchers improved transparency by providing supplemental information, such as visualization or explanation, to help decipher the existing results. Typical examples are post-hoc explanations that help users to understand the information provided by algorithms ~\cite{costa:18automatic, park:17uniwalk}. Facing with transparent information provide, users trust the system more and become more satisfied about their decisions ~\cite{cramer:08effects}. In this paper, we mainly focus on the second form of transparency (``what'') and explore how to use it to increase people's awareness of the self-selection bias in user-generated data.

Transparent design of data plays a vital role in users' decision-making in the area of visualization  ~\cite{Elis19:DataTransparency}. Similar designs for users to track and analyze data include the Google Dashboard ~\cite{Google:09dashboard} and  Mozilla’s Lightbeam ~\cite{Mozilla:13Lightbeam}. 
Similar works have explored ways to give users back control of their data through tools that enhance transparency, as summarized in ~\cite{janic:13transparency}. 
There also exist many works that leverage visualization to provide users with transparency about particular aspects of the data in different scenarios ~\cite{Julio15Transparency, ghoniem:04comparison, kolter:10visualizing}. For instance, Zavou et al. informed users via a ``chord diagram'' visualization that helps users understand how their data is treated by third-part cloud-hosted services ~\cite{zavou:13cloudopsy}.

In addition to enhancing data understanding and informing user decisions, transparency is also beneficial for \textbf{raising users’ awareness} of the hidden patterns behind data. 
A previous work pointed out a critical need for increased data transparency because the layperson might not be fully aware of how data aggregation was done by third-party services ~\cite{rader:14awareness}. Some other works emphasized that users should be aware of how systems or algorithms process user data and called for higher transparency ~\cite{Emilee18:Transparency, eslami2017careful}.
Transparent features were also used to improve users' awareness of their behaviors ~\cite{STEVENS:18Frame} and personal privacy issues ~\cite{Ebert:21BolderIB, kani:11increasing}. Visualization can be used to promote transparency and raise people's awareness of biases for informed decisions. The latest work by Arpit et al. proposed to raise users' awareness of their biased behaviors during data analysis for decision-making by showing the interaction traces ~\cite{narechania2021lumos}. 

In this paper, we set the goal of raising people's awareness of the self-selection bias in user ratings and reviews for informed decisions. We propose a \qian{bias-aware} design based on user ratings to reduce the impact of self-selection biases on people's decisions.

\section{Background} \label{background}

In this section, we first introduce the definition and \minor{the} threat of self-selection bias -- the focus of this paper -- in user ratings and reviews. Then, we propose to help raise people's awareness of this type of bias when referring to user-generated content by making information transparent. 

\subsection{Self-selection Bias on the Web} ~\label{self-selection bias}
In this work, we target the self-selection bias in user-generated data: \textbf{people with extreme experience (i.e., positive or negative experience) are more likely to give their feedback online than those who have a moderate experience} ~\cite{Bhole:17Self-SBias, li:08self}. 
Online ratings and reviews widely suffer from the self-selection bias ~\cite{aral:14problem, baeza:18bias}
and thus often fail to provide information about the quality of products/services that represents the general opinions of the entire user base~\cite{li:08self, de2016navigating}.

The self-selection bias is mainly caused by people's self-selected behaviors ~\cite{li:08self, Bhole:17Self-SBias} that can be unconscious or triggered by people's fast, instinctive processing system ~\cite{kahneman2011thinking, evans2009two}. For example, people could self-select to report or not report based on their experience (extreme or not extreme), and choose to merely report the aspects that leave the strongest impression.  
Such kind of self-selection behavior is part of human nature ~\cite{Demartini2021ManageBias}, thus, the resulting bias in user-generated data is hard to prevent and remove completely by algorithms or data management ~\cite{hettiachchi2021proceedings}. 

Furthermore, later users -- who refer to the ratings/reviews -- may found their personal experiences with the products/services inconsistent with their expectations established based on existing user reviews, as the biased feedback might not give a complete, up-to-date picture. 
The big expectation-experience disparity may cause a vicious circle of extreme feedback online, which hinders people's decision-making process ~\cite{Askalidis:17overcoming}.

Hence, it is critical to reduce the chance of people being \textit{``victims''} of the self-selection bias when referring to online ratings/reviews for their decisions. Raising people's awareness of the bias can guide us toward a solution to deal with biases in user-generated data ~\cite{baeza:18bias, Demartini2021ManageBias, hettiachchi2021proceedings}. In the next subsection, we introduce how we raise people's awareness of the bias by making them realize that certain information is over-represented in the sample of user reviews.

\subsection{Raising Awareness of Bias by Making Information Transparent}
We aimed to raise people's awareness of the self-selection bias by making three types of information that potentially reflect self-selection bias transparent in user ratings/reviews. 
We distilled three types of information by reviewing the literature ~\cite{li:08self, Askalidis:17overcoming, gong2015characterizing, baeza:18bias, Bhole:17Self-SBias,Ali:17Emotion, H:20Representative, halevy:19ubiquity}: 
(\textbf{I1}) the distribution of \textbf{reviewers} with difference experience who choose to share their opinions ~\cite{baeza:18bias, gong2015characterizing}, (\textbf{I2}) the distribution of \textbf{emotion extremity} ~\cite{Bhole:17Self-SBias, Askalidis:17overcoming, H:20Representative}, and (\textbf{I3}) the distribution of \textbf{reported aspect(s)} in user reviews ~\cite{li:08self, halevy:19ubiquity}. 
We introduce these pieces of information and why they need to be presented to users in details below.
\begin{itemize}
    \item We concerned the \textbf{composition of reviewers (feedback providers)} because those who leave comments online often cannot represent the entire population of people who have tried the products/services ~\cite{gong2015characterizing}. There exists the \textit{``silent majority''} who select not to give ratings or reviews on the Internet ~\cite{baeza:18bias}. Thus, we need to keep readers of online ratings/reviews informed of the characteristics of the reviewers.
    \item We cared about the \textbf{emotion extremity} of feedback because people tend to speak out when they have very strong feelings about something, which only constitutes a relatively small portion of the experiences of the majority ~\cite{Bhole:17Self-SBias, li:08self}. Thus, it is necessary to make users aware of the polarity and intensity of emotions behind the user ratings/reviews. 
    \item We considered the \textbf{reported aspect(s)} of user reviews because people can self-select to give feedback based on their own preferences ~\cite{halevy:19ubiquity}. They can be more sensitive to or impressed by one or several specific aspects of a product/service and write reviews (or give ratings) accordingly rather than equally weighting all aspects ~\cite{li:08self}. Showing the categories of aspect(s) that formed the basis of current user ratings/reviews could facilitate new users weight the feedback based on their own needs and interests.
\end{itemize}





Transparency can be an efficient way to raise people's awareness of the biased information from the perspective of how users perceive user ratings and reviews ~\cite{Demartini2021ManageBias}.
Before considering how to make the information transparent to raise awareness of the bias, we need to know exactly (1) how people perceive the three types of information and (2) what are the specific facets they would like to view about these three kinds of information when making decisions online. 
We conducted a formative study to collect users' opinions for these questions to inform the \qian{bias-aware} design.

\section{Formative Study} \label{formative study}
The goal of the formative study is to explore how people perceive the three types of information, and what exact information they care about when referring user ratings and reviews. 
We selected the hotel booking scenario among various scenarios because, comparing with other scenarios (e.g., watching a movie), people make more careful decisions in booking hotels since they pay relatively more money and time to buy and experience the service ~\cite{lin2009perceived}.

\subsection{Procedure}
We conducted an online survey with 206 participants who had booked a hotel online in the past two years. We performed the survey via Prolific\footnote{\url{https://www.prolific.co/}}, which is a crowd-sourcing platform designed specifically for online academic research. 
Prolific has been shown to provide better quality data from diverse participants than other platforms ~\cite{Peer:17Prolific}. 

We firstly set up a screening session to find participants who have experience in booking hotels online in the last two years. 
There are 273 of 400 participants who passed the screening session and 219 of them chose to participate in the formal survey. We collected 206 valid responses after excluding the participants who finished the survey in an extremely short time (<2 minutes). 

The content of the survey can be divided into three parts. 
\begin{enumerate}
    \item Participants first answered the questions about demographics and their habits of booking hotels, such as reasons of booking hotels, platforms they used, etc. We collected their habits as the guide for the later development of prototype systems with the \qian{bias-aware} design. 
    \item  Then, we collected participants' perceptions of the three types of information \textbf{(I1, I2, I3)} from two perspectives: how do they perceive the information and to what extent do they make decisions based on the information when booking a hotel. 
    To do so, we provided statements for each type of information and let participants to rate their agreement or disagreement on a scale of 1 to 7 (from strongly disagree to strongly agree). \\
    For example, the statements for the \textbf{extremity} of emotion (\textbf{I2}) are expressed like: \textit{``The extreme positive ratings/reviews are accurate''}, \textit{``The extremely positive ratings/reviews affect your likelihood of choosing a hotel''}, etc. The word \textit{``positive''} can be replaced by \textit{``negative''} or \textit{``moderate''} for different statements. To get the comprehensive perceptions, we also provide statements using different words beyond \textit{``accurate''}, such as \textit{``trustworthy''}, \textit{``reflect people's opinions''}, \textit{``reflect the quality of hotels''} or \textit{``reflect different aspects of hotels''}.
    \item In the third part, we collected the specific information participants care about when booking hotels online. We set questions like \textit{``How did you filter reviews when you book a hotel online?''}, \textit{``What kind of information of the reviewers you would like to view?''}, etc. 
\end{enumerate}
All questions are compulsory for participants and they need to return a complete code of Prolific after finishing the survey. 

\subsection{Participants}
We controlled the acceptance rate (>95\%) of all participants in Prolific.  
Each participant was paid \$2 for the survey that took eight minutes on average ($MAX = 11, MIN = 6$). 
These participants are all US-based (81 females, 122 males, 3 preferred not to disclose) with a mean age of 34 years ($MAX = 19, MIN = 71$). They came from different areas of the US (West 29.61\%, Northeast 23.79\%, Southeast 22.33\%, Midwest 15.53\%, Southwest 8.74\%). 
52.43\% of them were full-time employed, 17.96\% were students, 10.19\% were employed part-time, 6.80\% were unemployed, 4.85\% were self-employed, 2.43\% were retired, and 5.34\% selected ``other''.
Participants selected their highest educational level from the following options: high school degree (7.77\%), the college without a degree (17.96\%), associate degree (7.28\%), bachelor's degree (41.26\%), master's degree (20.39\%), professional degree (2.43\%), and doctorate (2.91\%).

\subsection{Findings} \label{findings}
\subsubsection{People pay attention to the background information of reviewers and mainly focus on reviewers' rating experience.} 
We found 58.77\% participants reported that they cared about the information of reviewers and 40.53\% participants considered the information of reviewers that affects their decisions when reading user reviews online. 
Among various kinds of information about the reviewers (e.g., age, gender), 61.92\% participants thought the rating experience of reviewers is critical for their decision-making online. The rating experiences include (1) the number of reviews written by a reviewer (31.07\%), and (2) the number of \textit{``helpful votes''} received by a reviewer (30.85\%).

\subsubsection{People's decisions are more likely to be affected by extreme positive or negative feedback than moderate ones.}
Participants thought the extreme positive ($MEAN = 5.67, SD = 1.16$) and negative ($MEAN = 5.5, SD = 1.22$) ratings and reviews affect their decisions more than the moderate ones ($MEAN = 4.99, SD = 1.42$). Thus, their decisions were influenced more by extreme positive or negative feedback.
Also, participants thought the emotions expressed at different extreme levels in user ratings/reviews are trustworthy that can reflect the overall opinions of users, as well as the quality of hotels.

\subsubsection{People care about the reported aspects when referring user reviews.}
In addition to the above findings, we found that participants preferred to filter reviews by their preferred aspects (23.39\%) and would like to read user reviews that have a photo so that they can check the detailed aspects of a hotel (27.58\%). People's preferred aspects of hotels are diverse (e.g., food or facilities), and they may associate the ratings they see with the aspects they care about, without being prompted for additional information ~\cite{halevy:19ubiquity}.

In conclusion, we found people mainly care about the reviewers' rating experience, including the number of written reviews and votes obtained, extreme feedback with positive and negative emotions, and the specific aspects reported in user reviews, and these kind\minor{s} of information indeed affect their decision-making when referring user ratings/reviews.
\section{User-centered Bias-Aware Design} 
To raise people's awareness of the self-selection bias in user ratings/reviews, we provided a \qian{bias-aware} design based on existing rating bar charts. In this section, we first describe our design requirements and then offer three alternative designs accordingly. 
We present the final design we chose according to the insights obtained from the pilot study, as well as the details refined in the design based on users' feedback.

\subsection{Design Requirements} \label{DRs}
\textbf{\textit{DR1: Disclose the distribution of the information behind user ratings.}} The current design of user ratings online merely provides the average rating and the number of different ratings (e.g., number of 1-5 stars). Therefore, it is hard for people to realize the self-selection bias hidden under user ratings and reviews with the current design. 
According to Section \ref{background} and Section \ref{formative study}, we aimed to disclose the distribution of reviewers' rating experience \textbf{(I1)}, the extreme level of emotions \textbf{(I2)}, and different reported aspects \textbf{(I3)} in user reviews to raise people's awareness of the self-selection bias. We discuss the details on how to extract these three types of information in Section ~\ref{extractI}.
Making people informed of how these aspects are distributed in user reviews can reduce the influence of bias on their decisions.

\textbf{\textit{DR2: Use simple and common visualizations for lay individuals to support their decisions.}} The target audience of our work is people who refer to user ratings/reviews online to make decisions, and they usually have no expertise in data visualization. Hence, the design should be simple and make the best use of common visual representations that people commonly encounter in their daily lives. Additionally, we considered the practicality and generality of the \qian{bias-aware} design by discussing with two visualization experts, as it aims to support people's daily online decisions. 
Therefore, we proposed three design alternatives based on the conventional design of user ratings (i.e., bar chart) rather than creating an entirely new design ~\cite{peck2019data}. 

\textbf{\textit{DR3: Enable filtering functions with the transparent information for exploring user reviews.}} When making purchase decisions, people spend the majority of their time exploring user reviews to mine suitable opinions for their personal decisions ~\cite{Xiaolan:20ExtremeReader}. According to the formative study, the information we aimed to disclose in the \qian{bias-aware} design is also critical for people's decisions. Thus, providing users with filtering functions based on transparent information is necessary to facilitate efficient decision-making.

\subsection{Pilot Study of Three Alternative Designs} \label{Alternatives}
\subsubsection{Three Alternatives}
\begin{figure}
\centering
  \includegraphics[width=\columnwidth]{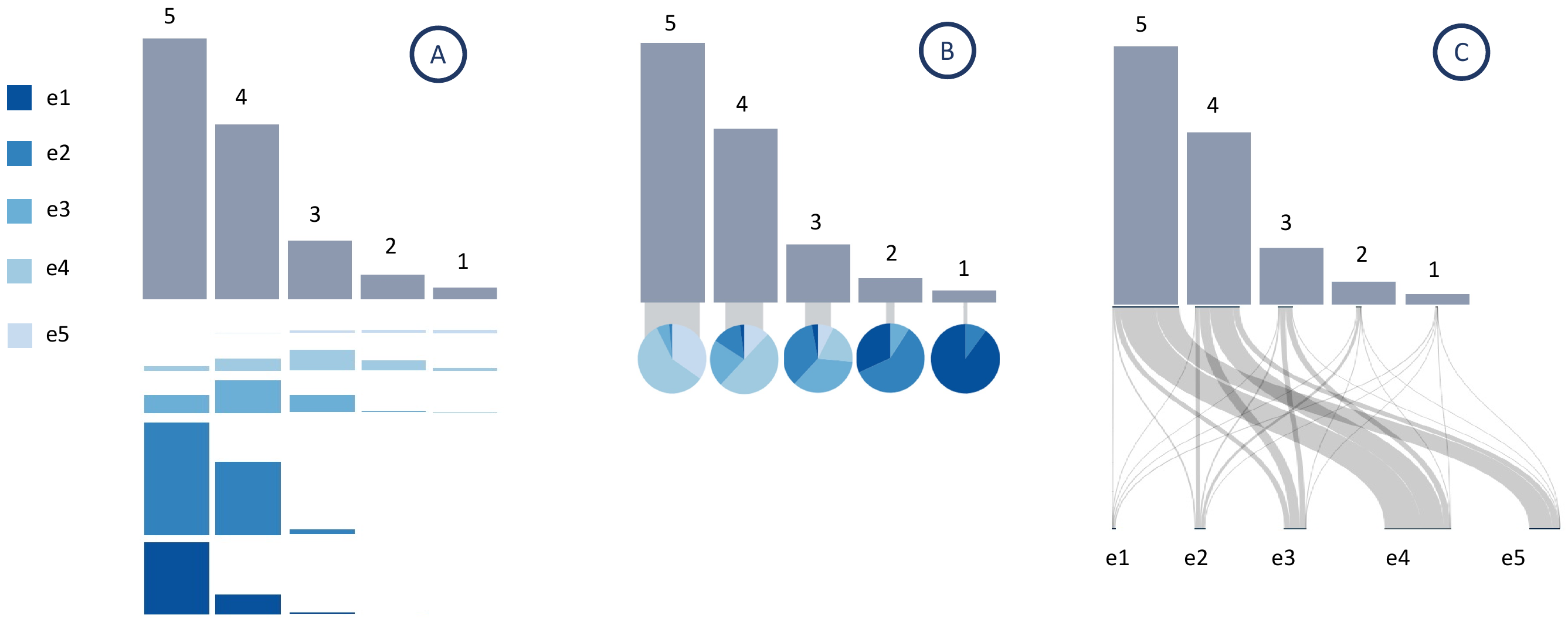}
  \caption{The alternative designs for the user ratings, where the proportions of information are encoded by: (A) a stacked bar chart design, (B) a pie chart design, and (C) a Sankey chart design.} 
  \label{fig:AlternativeDesign}
\end{figure}

To explore how to effectively show the distribution of information with user ratings \textbf{(DR1)}, we proposed three alternatives (Fig. ~\ref{fig:AlternativeDesign}). 
We first collected various visualizations that display data distributions based on the literature ~\cite{munzner2014visualization}, as well as the Internet\footnote{\url{https://datavizcatalogue.com/search/distribution.html}}. 
Then, we discussed with two visualization experts for two rounds to select suitable visual representations that can be understood by ordinary users while conveying the information effectively \textbf{(DR2)}. 
Finally, we selected the stacked bar chart (Fig. ~\ref{fig:AlternativeDesign} (A)), pie charts (Fig. ~\ref{fig:AlternativeDesign} (B)), and a Sankey chart (Fig. ~\ref{fig:AlternativeDesign} (C)) as alternatives to show the distribution behind user ratings. 
\qian{We used the real-world data of a hotel to demonstrate the alternatives. Details of extracting the transparent information and calculating the distribution of different categories are illustrated in Section ~\ref{extractI}. 

The three alternatives encode the same data derived from the user ratings and reviews of the example hotel. Since there are different types of transparency information to show (e.g., reviewer demographics or review sentiment), we use abstract variable labels (i.e., \eone\hspace{0.8 mm}e1 -- \efive\hspace{0.8 mm}e5 in the legend on the right of Fig. ~\ref{fig:AlternativeDesign}) to represent the different categories under each type of transparent information. 
For example, if the information currently encoded in the design alternatives is the emotion extremity (I2) of the user ratings as recognized by automatic algorithms, then \eone\hspace{0.8 mm}e1 denotes the category of extreme positive emotion while \efive\hspace{0.8 mm}e5 corresponds to the extreme negative emotion category. 
Based on the legend, in Fig. ~\ref{fig:AlternativeDesign} (A), the distribution of \eone\hspace{0.8 mm}e1 -- \efive\hspace{0.8 mm}e5 categories, regardless of data behind, is encoded by the length of stacked bars displayed under the user rating bars. The second alternative (Fig. ~\ref{fig:AlternativeDesign} (B)) shows the distribution of emotion (I2) categories by a pie chart under each rating bar, and the third alternative in Fig. ~\ref{fig:AlternativeDesign} (C) uses the Sankey chart, which encodes the distribution by the thickness of the branches. 

}



\subsubsection{Pilot Study}
To evaluate the three alternatives and find the most appropriate design, we conducted a pilot study in a one-to-one interview manner with 12 participants (six females, average age = 24), recruited via university emails. \qian{They had different backgrounds, including computer science (6/12), landscape architecture (1/12), finance (1/12), graphic design (2/12) and online education (2/12).} 

We started the pilot study with a brief introduction to the background of our work and the \qian{bias-aware design with three types of information}. Then we asked each participant to choose the most appropriate design and give reasons. We also asked participants how to interpret the additional information of each design by letting them imagine using these designs for booking a hotel online. We obtained participants' choices and feedback on each alternative during the study. One author analyzed the results independently and discussed them with the other three authors to identify user choices and needs.

\subsubsection{Result}
The second design in Fig. ~\ref{fig:AlternativeDesign} (B) got the most votes (8 out of 12) among the alternatives, as it was simpler, space-saving and easier to understand than others reported by participants. We analyze the reasons of participants' selection in this part and we will introduce the details of the design (B) in next subsection ~\ref{design}.

During the study, we observed that people interpreted information in the designs from two aspects. One is the distribution of information \qian{(i.e., \eone\hspace{0.8 mm}e1 -- \efive\hspace{0.8 mm}e5)} aligned with each rating bar, and the other is the distribution of one category \qian{(e.g., \eone\hspace{0.8 mm}e1)} across different rating bars (i.e., 1-5 points). 
We called these two aspects \textbf{``vertical''} and \textbf{``horizontal''} information acquisition, as the former focuses on the information under one rating bar while the latter spotlights one category across all different ratings.

Participants gave their reasons for selection according to these two aspects. 
For example, although the stacked bar chart (Fig. 2 (A)) provides an intuitive way for users to compare different categories (i.e., e1-e5) across different ratings \textit{``horizontally''}, it is hard for people to check the data \textit{``vertically''} under one rating, especially for the 1-star bar. 
Participants (7 out of 12) said they focused more on the 1-star bar than the other ratings (2–5 stars) when they imagined using the design for online decision-making. However, the 1-star bar in Fig. ~\ref{fig:AlternativeDesign} (A) is too thin for them to view and interact with. In addition, there are five participants reporting that the design in Fig. ~\ref{fig:AlternativeDesign} (A) has poor scalability when the number of categories increases as it will take too much space to show the stacked bars under the user rating.
For the alternative in Fig. ~\ref{fig:AlternativeDesign} (C), nine participants reported that it has too many lines, which makes their interpretation process difficult. Even if the design clearly shows the proportion of different categories under a rating bar by the thickness of branches, the intersection of these lines makes it difficult for people to compare and analyze data \textit{``vertically''} and \textit{``horizontally''}. 
Moreover, four participants said that the alternative in Fig. ~\ref{fig:AlternativeDesign} (C) is a bit complex for them as they had not used the Sankey chart previously (DR2).

Therefore, we added an additional forth design requirement \textbf{DR4} based on participant's feedback. \textbf{\textit{DR4: Support interactive data comparison among different categories (\textit{vertically}) and user ratings (\textit{horizontally}).}} As users need to view and compare the transparent information with user rating bars from these two aspects, it is necessary to provide an efficient way for people to access the distribution \textit{vertically} and \textit{horizontally}. 
We further improved the design in Fig. ~\ref{fig:AlternativeDesign} (B), which is selected by most participants, based on the \textbf{DR4}. 
We detail how we improve the design with interactive functions in the next subsection ~\ref{design}.

\begin{figure}
\centering
  \includegraphics[width=\columnwidth]{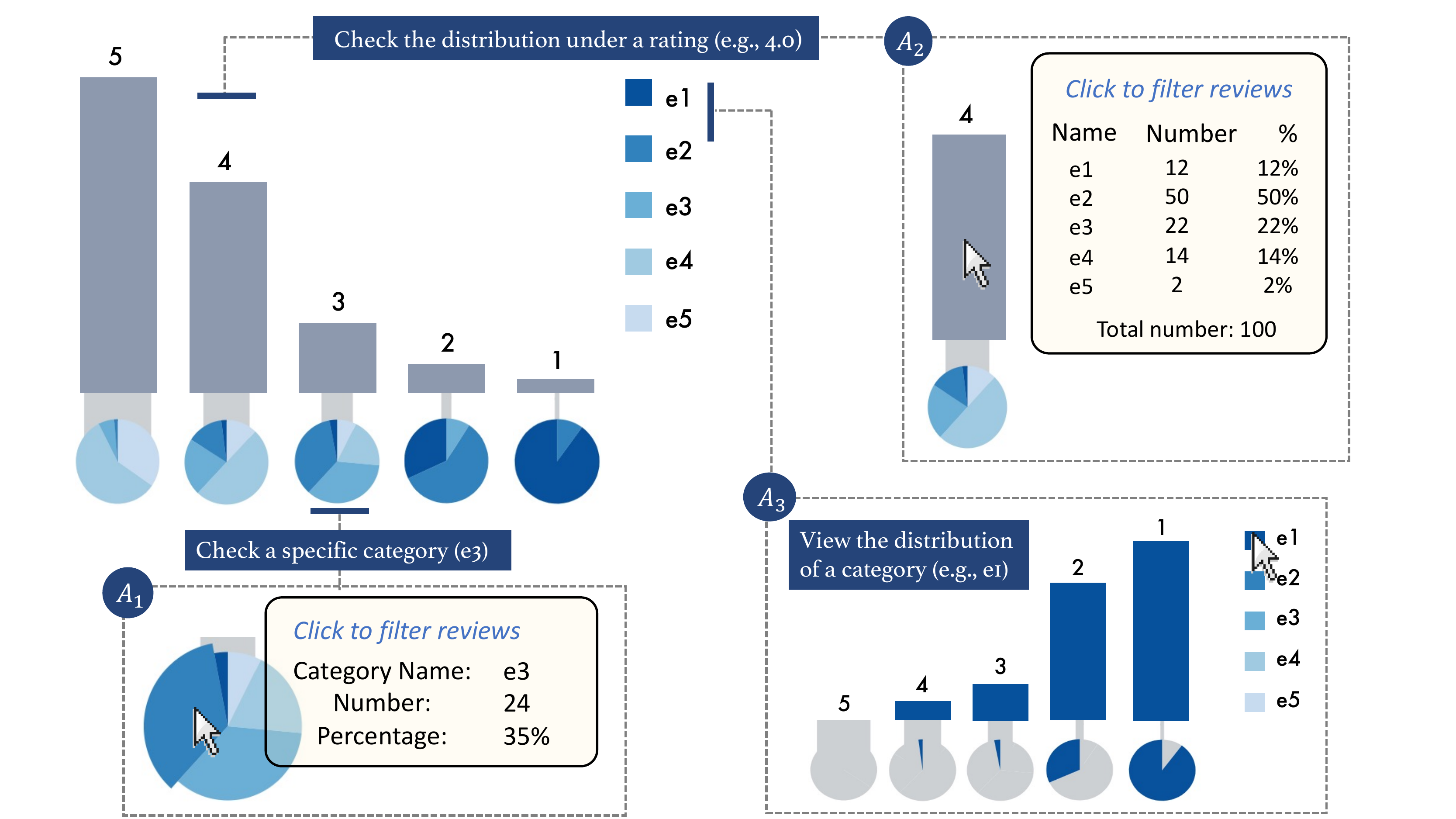}
  \caption{The \qian{bias-aware} design and with three interactive functions. The distributions are encoded in pie charts for each rating bar. (A1) shows the hover effect on the pie chart; (A2) displays the hover effect on a rating bar, and (A3) is the result triggered by hovering on the legend rectangles. The user rating bars in (A3) can vary according to the selected category (e.g., e1) by hovering over a category in the legend. } 
  \label{fig:Design}
\end{figure}
\subsection{The Final \qian{Bias-aware} Design and Interactive Functions} \label{design}
As shown in Fig. ~\ref{fig:Design}, we present the final design of the transparent information with user ratings. The design combines pie charts with user ratings that clearly show the distribution of information \qian{(e.g., \eone\hspace{0.8 mm}e1 -- \efive\hspace{0.8 mm}e5)} \textit{vertically} in user reviews behind each bar \textbf{(DR1)}. It shows the categories across user ratings \textit{horizontally}, leveraging the interactive design in Fig. ~\ref{fig:Design}(A3).
The pie chart is a conventional chart showing distributed data, which is familiar to people in daily life \textbf{(DR2)}. The legend on the right-hand side of the design shows the color scale associated with each category \qian{(i.e., \eone\hspace{0.8 mm}e1 -- \efive\hspace{0.8 mm}e5)} of the information (i.e., the extremity of emotion). To coordinate the proportions of elements in the design and improve space utilization, we keep all pie charts the same size and use the thickness of the (light grey) lines linking pies and bars to represent the number of ratings/reviews for each pie. 

We designed three interactive functions for the design in Fig. ~\ref{fig:Design} according to the participants' requirements in the pilot study. These functions could help users compare the detailed information with the design \textbf{(DR4)} while offering filtering functions to filter user reviews in a specific aspect \textbf{(DR3)}. 
In Fig. ~\ref{fig:Design} (A1), we added a zoom effect on each sector of a pie chart so that users can easily hover over them to view the details of each category in a floating text box and filter user reviews by clicking a sector. 
Similarly, in Fig. ~\ref{fig:Design} (A2), users can check the information of all categories under a rating by hovering on the bar and clicking to filter reviews by a rating. 
To meet the \textbf{DR4}, we designed a mouse-over triggered animation with the legend. By hovering on the legend, users can view the distribution across different ratings \textit{horizontally} related to one category (Fig. ~\ref{fig:Design} (A3)). The corresponding sectors of the selected category \qian{(e.g., \eone\hspace{0.8 mm}e1)} in different pie charts will also be highlighted. In this way, users can easily check the distribution of one category across different user ratings.
We added a smooth transition animation to make the changes between bar charts more natural. Moreover, users can also filter reviews by clicking the categories in the legend, such as filtering reviews that express extreme negative emotions \qian{(i.e., clicking \eone\hspace{0.8 mm}e1)}.

\section{Prototype System}  \label{system}
To evaluate the proposed design with users, we built a prototype system with transparent design for booking hotels. We firstly collected the data of real hotels from one of the most popular platforms, Tripadvisor\footnote{\url{https://www.tripadvisor.com/}}. Then, we extracted the specific aspects of the three kinds of information from the data automatically. Finally, we integrated the \qian{bias-aware} design with data in a prototype system simulating the hotel-booking scenario online.

\subsection{Data Collection} \label{datacollect}
To evaluate our design effectively, we aimed to collect data from representative hotels online, which could mirror the opinions of ordinary users.
Thus, we decided to select hotels from big cities in the world. 
Considering that participants of the formative study were all US-based, we retained this character in the user study. Hence, to reduce the impact of their previous impressions, we collected data from non-US cities that participants might not be familiar with. 
Two authors explored and compared the hotels' data in different cities. 
\qian{Considering that the COVID-19 may influence the user ratings and reviews of hotels online, we only collected user ratings/reviews in the time interval of 30 June, 2016 to 31 January, 2020 (before COVID-19 became International Concern~\footnote{The World Health Organization declared the outbreak a Public Health Emergency of International Concern on 30 January 2020, and a pandemic on 11 March 2020.}). Indeed, the pandemic is likely to change demographics of travelers, their emotions, and the aspects they concern about a hotel. It would be interesting to explore and compare the changes of content and the potential biases in the content of user feedback before and after COVID-19. We leave this topic as part of our future work, because COVID-19 made travel extremely difficult, if not impossible, in most parts of the world when we conducted the experiments.}
Finally, we chose London as it had the most hotels as well as user reviews. 

We initially crawled 976 hotels' pages under the tag \textit{``London''} on Tripadvisor. To narrow down the scope and select representative hotels, we narrowed down the scope of hotels to 3-star hotels labelled by Tripadvisor as hotel-class because there are more 3-star hotels (40.7\%) than others. 
These hotels had a price range from \$40 to \$200, with \$88 as the median price, and we found that hotels with price ranges from \$82 to \$105 contained more than 80\% of the user ratings/reviews within the setting time interval. Therefore, we selected hotels based on this price range and filtered out the hotels with no feedback for the last six months. Finally, we got 57 hotels and crawled the data of these hotels, including the hotels' names, price, user ratings, reviews, etc. 
By controlling these variables, we could select the representative hotels and let users focus more on the user feedback to evaluate the \qian{bias-aware} design effectively.

\begin{figure}
\centering
  \includegraphics[width=\columnwidth]{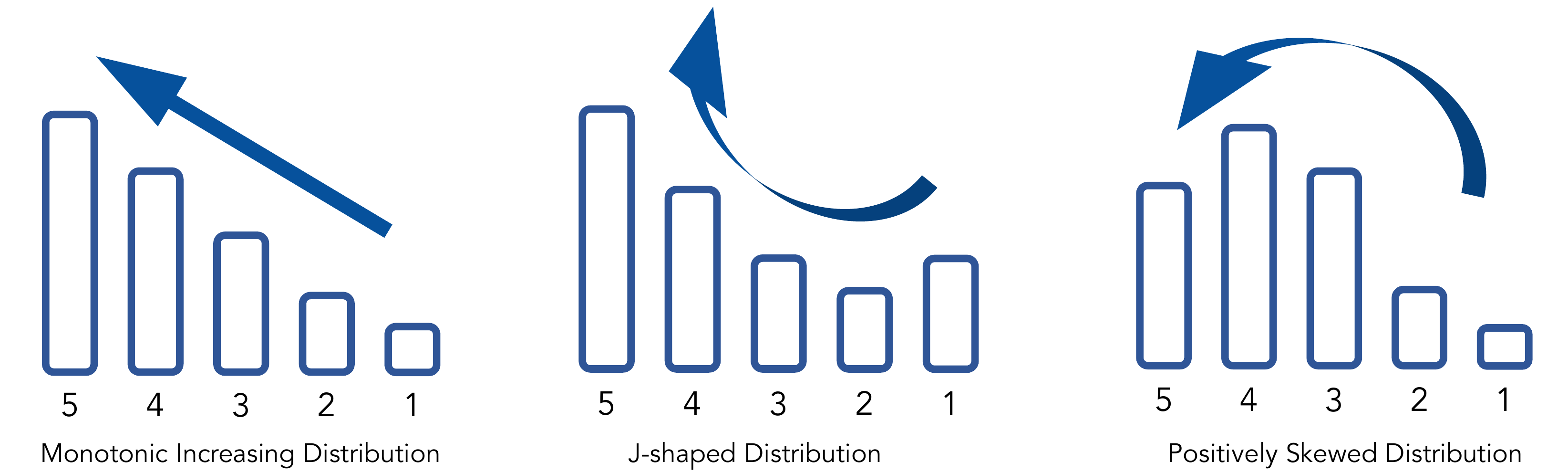}
  \caption{Three kinds of distribution shapes for user ratings, observed from real data on Tripadvisor.} 
  \label{fig:Distribution}
\end{figure}
\subsection{Data Distribution} \label{3-distributions}
After selecting the representative hotels by control variables, such as prices and the hotel-class, we explored the distribution shapes of user ratings in these hotels as the distribution can be affected by the self-selection bias proven by previous works ~\cite{Verena:18Extreme, hu:09overcoming}. 
We analyzed the user rating data of the 57 hotels and summarized three kinds of distribution in Fig. ~\ref{fig:Distribution}. 

These three distributions include (1) the monotonic increasing distribution, (2) the J-shaped distribution, and (3) the positively skewed distribution, which are observed from the selected hotels' ratings on Tripadvisor. 
(1) The monotonically increasing distribution means that the shape of the user rating bars is monotonically increasing from 1 point to 5 points. 
(2) The J-shaped distribution is the typical distribution caused by self-selection bias as it has an extreme distribution with more 1-point and 5-point ratings ~\cite{hu:09overcoming, Bhole:17Self-SBias}. 
(3) The positively skewed distribution has more middle points (from 2 to 4 points) with fewer 5-point ratings compared to the other two distributions. This distribution is close to real-world data under lab conditions ~\cite{lim:17mitigating}. 

We randomly choose 15 hotels from 57 hotels with five for each distribution shape, considering that the distribution shape may affect people's perceptions of hotels, and their decisions and people usually narrow down to no more than 15 hotels for decision-making based on the pilot study.
We further extracted and analyzed the user ratings and reviews for the selected 15 hotels, which is illustrated in the next subsection.


\subsection{Data Processing} \label{extractI}
The data processing step consists of three parts corresponding to the three kinds of information mentioned in Section ~\ref{formative study}. We collected 5940 user ratings and reviews from the 15 hotels (MEAN = 368, MAX = 397, MIN = 320) in total between 30 June 2016 to 31 January 2020. 
We first obtained the reviewers' information (I1) based on the \textit{reviewer badge} on Tripadvisor. Then we got the information about emotion (I2) and reported aspects (I3) using two automatic algorithms in the natural language processing (NLP) domain, including sentiment analysis and topic extraction. Based on these approaches, we mapped the information (I1-I3) into the \qian{bias-aware} design.

\subsubsection{\textbf{I1: Reviewers' Rating Experience}}
We measured the reviewers' rating experience by separately calculating the number of total reviews they gave (I1-1) and the number of ``helpful'' votes they got (I1-2) on Tripadvisor. Then, we divided six categories for each kind of information referring to the reviewer badge on Tripadvisor. 
For example, we marked a reviewer as a \textit{``Top Reviewer''} when they have more than 100 reviews and annotated a reviewer as a \textit{``New Reviewer''} if he/she has only posted one review. For the second type (I1-2), referring to the classification of Tripadvisor, a reviewer with more than 100 ``helpful'' votes would be called a ``Top Contributor'' while a reviewer with 0 ``helpful'' votes is considered a ``New Contributor''.
Hence, we can label each review by its reviewers' rating experience in these two aspects (each has six categories) and calculate the distribution of user reviews under these categories associated with user ratings.


\subsubsection{\textbf{I2: Sentiment Analysis of User Reviews}}
To extract the emotional polarity from textual reviews, we performed sentiment analysis on user reviews with AllenNLP ~\cite{Gardner2017AllenNLP}. It provides an off-the-shelf, ready-to-use sentiment analysis library based on Roberta ~\cite{liu2019roberta} and the model was trained on the Stanford Sentiment Treebank ~\cite{socher2013recursive} with a test accuracy of 95.11\%. 
We divided the results into five categories ranging from $-1.0$ to $1.0$ evenly and used one of the categories, including \textit{``Positive Only''}, \textit{``Positive''}, \textit{``Neutral''}, \textit{``Negative''} and \textit{``Negative Only''}, to label the extremity of emotion in a user review. We obtained the distribution of the extremity of emotions associated with user ratings by counting the number of reviews in each category and their corresponding percentage.

\subsubsection{\textbf{I3: Reported Aspects of User Reviews}}
We used a topic extraction approach to extract keywords from user reviews as the reported aspects (I3). Firstly, we trained the topic extraction model using another hotel review dataset\footnote{\url{https://www.cs.cmu.edu/~jiweil/html/hotel-review.html}} which consists of 878,561 reviews from 4333 hotels on Tripadvisor. Then, we applied the KeyBERT ~\cite{grootendorst2020keybert} model to extract keywords from reviews based on the pretrained model. We keep the 300 most frequent keywords from the extracted keywords and convert each of them to a vector by looking up its word embedding vector in GloVe. Through clustering these 300 vectors with K-Means algorithm, we found nine clusters.

Three authors analyzed the resulting clusters and excluded two clusters after discussion, whose keywords are the dates of user review (e.g., March or 15th). We also merged two clusters that are similar and are related to the environment of hotels. 
Finally, we got six categories that described different aspects of hotels in user reviews (i.e., food, facilities, service, surrounding environment, travel purpose, and companions). As one review may contain multiple categories, we calculated the percentage of a category by adding up the total number of reviews in all categories.

\begin{figure}
\centering
  \includegraphics[width=\columnwidth]{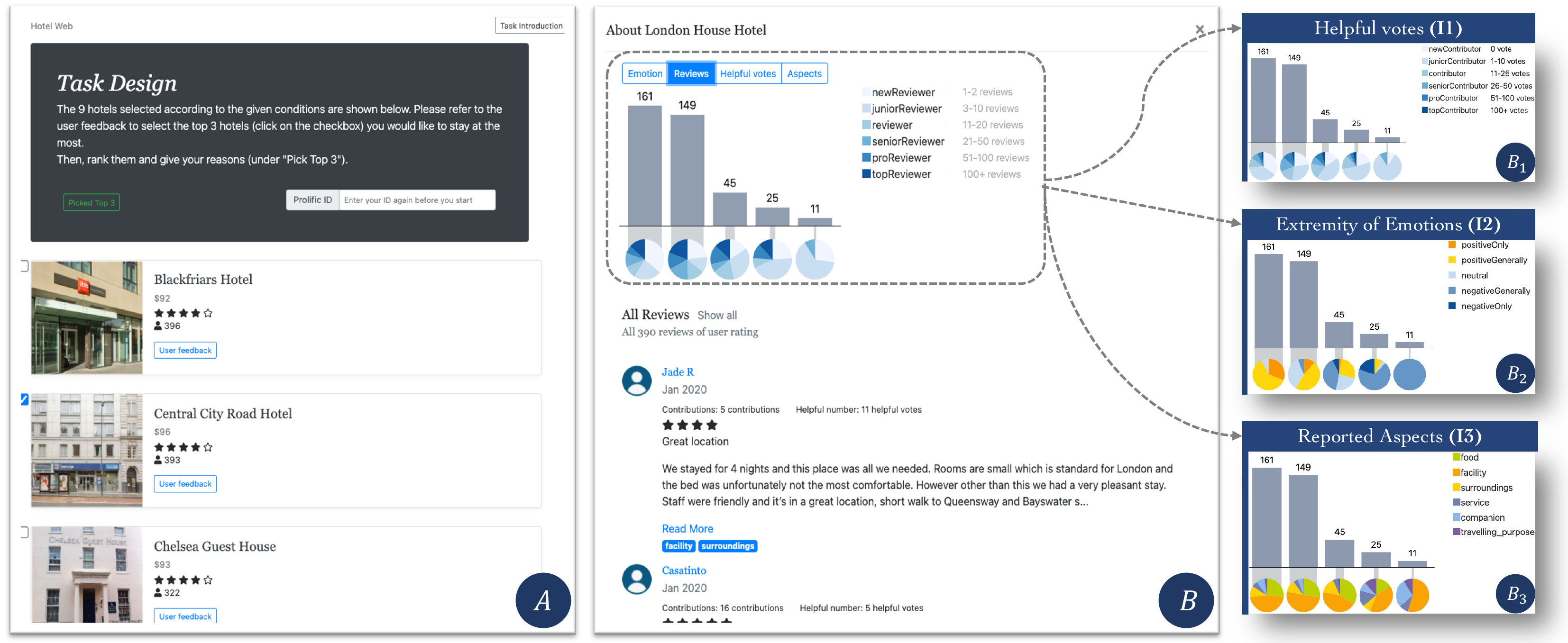}
  \caption{Two main interfaces of the prototype website. (A) shows the list of hotels provided for participants to select from; (B) displays a detailed page of one hotel, including the \qian{bias-aware} designs of user ratings. Users can switch the design with different types of information by clicking the navigation bar above.} 
  \label{fig:prototype}
\end{figure}
\subsection{A Hotel-booking Website with \qian{Bias-aware} Design}
To evaluate the design with ordinary users, we simulated a hotel booking scenario by building a prototype website integrating the \qian{bias-aware} design. To increase the scalability and user friendliness of the prototype, we implemented the basic functions of booking hotels and used a responsive framework that makes the interface adaptive to different screen sizes.

We first collected the interface design of many popular hotel booking websites on the Internet, such as Tripadvisor.com, Booking.com, Expedia.com, etc. We analyzed their interfaces and found that these websites use a similar framework of user interfaces, which shows the general information of each hotel in a card view and displays the detailed user reviews in a separate web page or a pop-up page. 
We adopted this design and integrated the \qian{bias-aware} design into the user rating position in the detailed page of each hotel. 

We show the two main interfaces of the prototype in Fig. ~\ref{fig:prototype}, in which (A) shows the homepage with the introduction of the user study and a list of hotels. By clicking the button in a card view, users can open a new page in a pop-up window that displays the corresponding user ratings and reviews of one hotel in Fig.~\ref{fig:prototype} (B). We display the \qian{bias-aware} design with a navigation bar above, which is used to switch between different information types (Fig.~\ref{fig:prototype} (B1)(B2)(B3)). There are four options in the navigation bar that include two types of rating experience of reviewers (i.e., written reviews (I1-1) and ``helpful'' votes (I1-2)), emotion (I2) and aspects (I3). Users can interact with the design to filter user reviews below or check the distribution of one category across all different user ratings, as illustrated in Section ~\ref{design}.

We display the user reviews below the \qian{bias-aware} design and show each review together with the user name, time, reviewer, user rating, and reported aspects, referring to the design of Tripadvisor. Users can scroll up and down to read user reviews in this page. To avoid gender bias, we replaced all the original reviewers' names with either abbreviations or random common names for both males and females.


\section{User Study} \label{studyDesign}
With institutional IRB approval, \qian{we conducted a between-subjects user study with 144 participants to evaluate the effectiveness of the bias-aware design.
Our goal is to answer the general research question through the evaluation: \textbf{(How) does the bias-aware design help people make informed decisions?} We further decomposed  the general research question into three subquestions.
\begin{itemize}
    \item \textbf{RQ1}: Does the bias-aware design raise people’s awareness of the self-selection bias compared to the baseline (the common design of user ratings)?
    \item \textbf{RQ2}: How do people use the bias-aware design to make decisions compared to the baseline?
    \item \textbf{RQ3}: How does the bias-aware design affect people’s decisions compared to the baseline, if at all?
\end{itemize}
The goal of \textbf{RQ1} is to test the design in raising people's awareness of the self-selection bias compared to the baseline. In addition, to gain insights into improving our design and offering implications for future work, we observed and compared users’ decision-making strategies in both conditions \textbf{(RQ2)}. We set \textbf{RQ3} to investigate the effect of the bias-aware design on users' final decisions, and further verified that enhancing people’s awareness of the self-selection bias could help people make more informed decisions.
}


\subsection{Participants}
We used the Prolific to conduct the user study online. To ensure that all participants had the experience of booking hotels online prior to our study, we set screening questions on the frequency and most recent activities of booking hotels. We initially recruited 144 participants and excluded the responses of eight participants who failed this quality control during the study. 

We obtained responses from 136 participants after the quality check, of whom 68 (33 females, 34 males, and one prefer not to say) completed the experiment with our proposed prototype system, and the other 68 (35 females and 33 males) used the baseline system. The distribution of our participants' education background is as follows: undergraduate (46.3\%), graduate (29.4\%), technical/community college (8.1\%), high school diploma (8.8\%), doctorate degree (5.9\%) and unknown (1.5\%). \qian{The average age of the participants was 31 (Max= 73, Min = 20, SD = 9.97, 95\% CI [29.14, 32.49]).}
Following the rules of the Prolific platform, we paid all participants at a rate of £7.60/hour. The actual amount for each individual depended on the task completion time and the Prolific settings according to our predefined time range (30 min).

\subsection{Experiment Setup and Task Flow} 
\begin{figure}
\centering
  \includegraphics[width=\columnwidth]{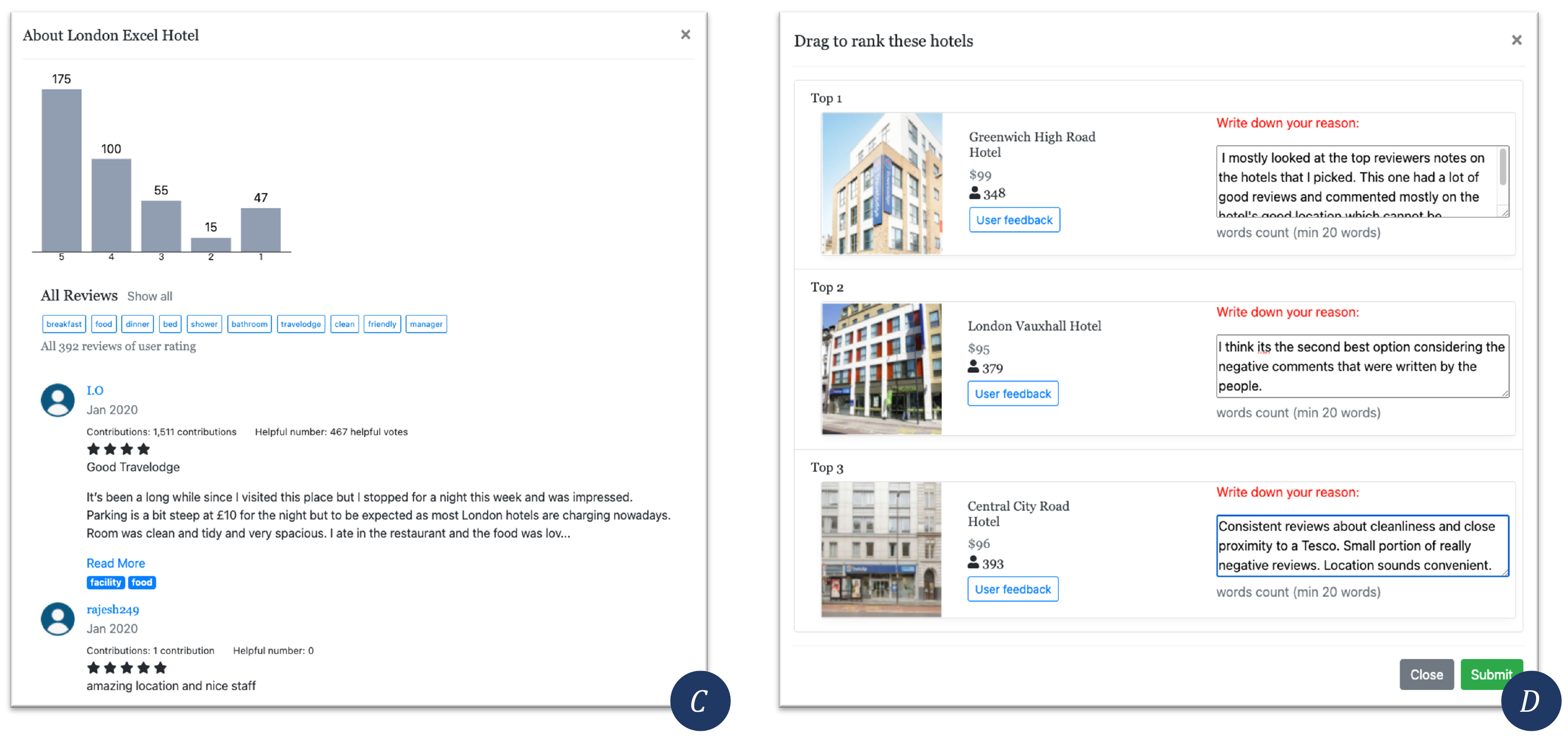}
  \caption{The interface provided by the baseline system is shown in (C). The interface in (D) is used to record the results of user selections and the reasons for uploading in both prototype and baseline systems.} 
  \label{fig:baseline&Task}
\end{figure}
\subsubsection{Baseline System}
For comparison, we created a baseline system by removing our visual design but keeping other hotel booking-related functions available in the prototype system. 
The baseline system shared the same home page (Fig. ~\ref{fig:prototype} (A)) with our proposed system, but it adopted a simpler visualization of user ratings in the detail page, as shown in Fig. ~\ref{fig:baseline&Task} (C), on which users could filter reviews by semantic tags or rating bars, like the common hotel booking sites used in daily lives. We hosted both systems on a web server that enables participants to gain access via a public link.

\subsubsection{Configuration of the Website}
To reduce the confounding effect of irrelevant variables in the study, we controlled the information presented to make participants perceive a hotel primarily by viewing user ratings and reviews. 
We preprocessed the information presented to participants on both websites, such as the price, number of user reviews, and user ratings of each hotel, within a narrow range, as described in Section ~\ref{datacollect}. 
During the study, the candidate hotels were presented in a randomized order for each participant to avoid anchoring bias. Moreover, we kept the page's functionality simple. We only used photos of the hotels' entrances or buildings to reduce any influence of photos on the participants' decisions.

\subsubsection{Task Flow} \label{taskflow}
We designed a task for user study that simulates the real-world online hotel booking experience. To complete the task, the participant had to go through the following three steps. 
\begin{enumerate}
    \item They first watched an introductory video\footnote{We upload the introduction video for the experiment in the supplementary materials} (1-2 minutes) about the usage and task of the experimental websites. Through the video, we let the participants imagine that they were required to choose a hotel for a vacation in London based on user feedback, while paying less attention to other information controlled in our study, like price or hotel class. 
    \item Then, participants browsed a pool of 15 candidate hotels (identical for the experimental and baseline conditions), including the user ratings and reviews in the given system and shortlisted the top three hotels they may want to make a reservation at. To compile the shortlist, the participants needed to specify three hotels by checking the box next to the hotel name on the homepage (Fig. ~\ref{fig:prototype} (A)). The choice of hotel depended totally on the participant’s personal judgment. 
    \item After that, participants could rank the selected hotels by dragging the hotel up or down in a pop-up window (Fig.~\ref{fig:baseline&Task} (D)). They needed to justify their decisions in the text area of each hotel. Note that their draft answers in the pop-up window were automatically saved, so they could return to the previous page to view the details of a hotel at any time if needed. To conclude the task, the participants had to submit their final choices and feedback by clicking the \textit{Submit} button in Fig.\ref{fig:baseline&Task} (D). 
\end{enumerate}
These three steps allowed us to examine the effectiveness of our design and collect information about how people make decisions within the two systems. 

\subsection{Pilot Study}
Before launching the formal experiment, we conducted a pilot study to test our task design and configuration with eight participants (four females), who were undergraduate or postgraduate students recruited from a local university. \qian{Their ages ranged from 21 to 27 (Mean = 24, SD = 2.51, 95\% CI [21.90, 26.10]), and their backgrounds included computer science, finance, psychology, and biology.} These participants were randomly divided into two groups; half completed the task with our prototype, while the rest used the baseline system. After finishing the task, they filled out a post-study questionnaire and participated in a brief interview to share their study experiences. 

Based on the feedback from the pilot study, we adjusted several designs in both conditions. 
\qian{Seven out of eight participants (87.5\%) reported that they usually inspect and compare fewer than 10 hotels in detail in real life, because it became overwhelming and time-consuming beyond that number. Examining 15 hotels fully deviated from their common practice and they were not able to  remember and compare the information of all these hotels at once. }
Thus, we reduced the number of hotels in the candidate pool from 15 to 9 for our formal study by randomly removing two hotels from each distribution type mentioned in Fig. ~\ref{fig:Distribution}. Furthermore, as informed by the pilot study, we further streamlined our study procedure (presented in the next subsection) and enriched the details of the instruction videos. 

\subsection{Procedure} \label{studyprocedure}
In our formal study, 144 participants were randomly split between the baseline group and the experimental group. All participants first signed a consent form, agreeing to join the experiment and allowing us to collect basic information about them. We recorded their operations during our study and obtained their demographic data from the Prolific platform. 
Then, following the three steps presented in the task flow (Section ~\ref{taskflow}), each participant watched an introduction video, browsed the candidate hotels, specified the top three choices, wrote down the reasons for the choice, and submitted in the prototype system. 
After sending their selections and reasons, each participant filled out a post-study questionnaire, which assessed their awareness of potential bias and collected qualitative feedback on our design and systems \qian{(RQ1)}. The average time for the whole study was 34 minutes (baseline: 29 minutes, prototype: 39 minutes). \qian{We did not control the time of our user study because we aimed to simulate the real-world scenario of booking hotels so that the study results could truly reflect people’s decisions and their decision-making process. We concerned that strictly controlling the time may add stress to the participants and affect their behavior during the experiment and consequently the observed results.}

\subsection{Operational Data Collection and Analysis} \label{TaskResult}
\qian{
\paragraph{\textbf{General quality check.}} 
We carried out a general quality control by checking the usage time of each participant. The time started counting from when they opened our prototype and ended by submitting their feedback to the questionnaire. They can get the completion code and return it to Prolific after successfully finishing all questions in the questionnaire. Eight responses (out of 144) were removed due to the extremely short time recorded (i.e., less than 1 minute for each hotel on average, set based on the pilot study) or they did not return the completion code of Prolific. Finally, we got 136 valid responses to the questionnaire (68 responses for each condition).
}

\paragraph{\textbf{Quality check of interactive operations.}}
\qian{To analyze and compare users’ behaviors in both conditions, we implemented several event listening functions in our prototype to collect participants' operations during the study (RQ2).} We collected all records of participants' interactive behavior for browsing and selecting from nine candidate hotels. The behavior logs include the basic interactions of all participants, including clicking, hovering, and scrolling up or down in our system. We counted and averaged the number of user interactions, i.e., clicks and hovers, on each individual rating bar. The interactions recorded with the user rating bars in the prototype include those \qian{on} the pie chart linked to each bar (Fig.~\ref{fig:Design}). 

\qian{
To ensure the completeness of the data logs of participants’ operations (e.g., click, scroll, hover, etc.) on the study interface (RQ2), we defined a threshold of the minimum number of interactions required for each participant to check the information of all 9 hotels provided in the study. This threshold,  102 times of operation, is set based on the observation of pilot study. After scrutinizing the operation data using the  threshold, we identified 26 behavior logs (out of 136) that failed the quality check. 
We found that these participants all had high credit scores on Prolific and their operational logs are incomplete because of the lack of hovering operation records, which indicated that they might not fail the check on purpose. To figure out the reasons, we tried to get in touch with these 26 participants via the Prolific chat. Through multiple rounds of communication, we confirmed that these participants failed the quality check due to the compatibility issues occurring with older browsers. Therefore, we only removed their operation logs from the data analysis while keeping their feedback on the questionnaire. Finally, we retained 136 valid responses to the questionnaire and 110 behavior logs (53 with the baseline, 57 with our design) from participants. } 


\section{Results} 
\qian{In this section, we structure the results according to the research questions mentioned in Section ~\ref{studyDesign} (RQ1-RQ3). We first provide the results of the questionnaire reflecting participants’ awareness of self-selection bias when making decisions (RQ1). We next show and compare the strategies used by the participants to make decisions in the two conditions of the study (RQ2). We also summarize patterns in participants’ operational logs in both conditions to reflect how the experimental group  used our prototypes and compare their actions to those in the baseline condition (RQ2). Finally, we present and compare the final selections between the two conditions (RQ3).}

%

\qian{
\subsection{Raising Awareness of the Self-selection Bias (RQ1)}
We measured people’s awareness of the self-selection bias in user ratings and reviews by the post-study questionnaire. Note that the self-selection bias is implicit, which needs to be measured carefully without revealing the goal of our study (testing awareness) during the experiment. To effectively test whether the bias-aware design works or not, we instructed each participant to book a hotel using the given prototype system under a concrete scenario (going to London on vacation) as how they would do it in the real life. We followed a psychological method called implicit association test (IAT) ~\cite{nosek2005understanding} and adapted the questionnaire of a famous IAT project called Project Implicit\footnote{\url{https://implicit.harvard.edu/implicit/education.html}} to design our questions for accessing awareness.
}
IAT is widely used to assess people's implicit bias, stemming from attitudes or stereotypes that mostly occur outside of people's consciousness and control ~\cite{nosek2005understanding}. 

\begin{figure}
\centering
  \includegraphics[width=\columnwidth]{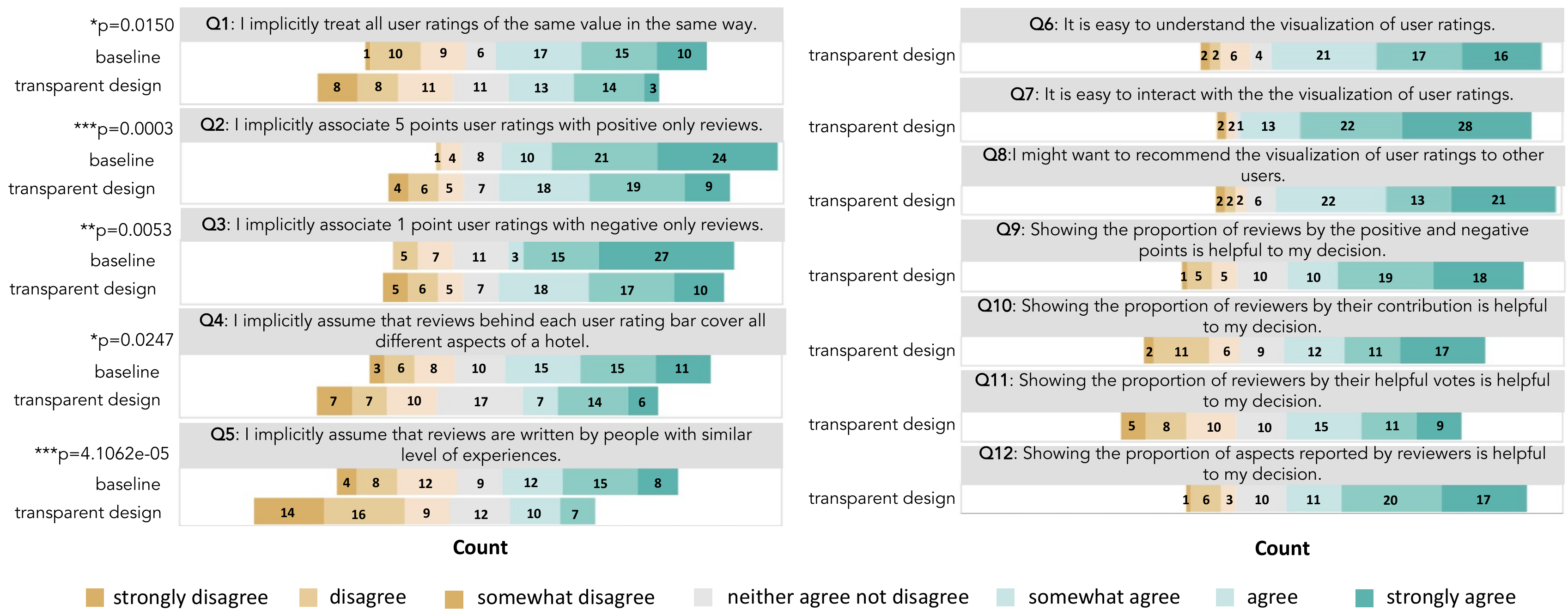}
  \caption{Participants' ratings of the awareness testing questions in post-questionnaire and the subjective response to our design in the experimental group.} 
  \label{fig:questionnaireRate}
\end{figure}

As shown in Fig. ~\ref{fig:questionnaireRate}, we present our five adapted IAT questions (Q1–Q5) with the corresponding participant ratings that measure their overall awareness (Q1) and the awareness of the three types of information (Q2-Q3: emotion [\textbf{I2}], Q4: aspects [\textbf{I3}], Q5: reviewers [\textbf{I1}]) caused by the self-selection bias. We used \qian{the Mann-Whitney test \cite{mcknight2010mann} to compare the result of both conditions and calculated the 95\% bootstrap confidence intervals for each question \cite{diciccio1996bootstrap}.} We display the responses of other post-study questions (Q6-Q12) that evaluate the information provided by the transparent design as well as its usability on the right side of Fig. ~\ref{fig:questionnaireRate}.

The results of Q1–Q5 show that people in the experimental condition disagreed more with the descriptions, which indicated that they were significantly more aware of the self-selection bias in user ratings and reviews. 
Q1 measures the general awareness of participants between two conditions, which indicates that people booked hotels with our design were more aware of the diverse information behind user ratings. 
Among these cases, participants were more aware of the diverse experience of reviewers behind user ratings (Q5, p-value < 0.001) more than other cases using the transparent design. 
This result echoes the last strategy used by participants when selecting hotels in Section ~\ref{strategy}.
For the responses of Q4, participants' awareness of the different aspects of a hotel behind user ratings (p-value = 0.0247) had the least difference between the two conditions. 
The reason is that participants using the baseline could also view the tags of diverse aspects for filtering reviews under the user rating as the ordinary hotel booking websites.
Q2 and Q3 evaluate the level of participants' awareness of the extreme emotions behind user ratings, including the positive and negative emotions. Participants who used the transparent design were relatively more aware of the negative emotions (Q3, p-value < 0.001) than positive ones (Q2, p-value < 0.01). 

We collected participants' feedback on the information and usability of our design (on the right side of Fig.~\ref{fig:questionnaireRate}) and found that our design was perceived to be easy to understand \qian{(Q6, Mean = 5.28, SD = 1.51, 95\% CI [4.92, 5.64]) and interact with (Q7, Mean = 5.96, SD = 1.30, 95\% CI [5.64, 4.26])}. 
In general, participants would like to recommend our design to other people \qian{(Q8, Mean = 5.46 , SD = 1.48, 95\% CI [5.12, 5.82])}. We also asked the participants who used the transparent design about the helpfulness of the information provided (Q9-Q12). 
Overall, participants agreed that showing the proportion of aspects \qian{(Q12, Mean = 5.24, SD = 1.63, 95\% CI [4.84, 5.61]) and emotions (Q9, Mean = 5.24, SD = 1.62, 95\% CI [4.86, 5.64])} behind user ratings is useful for their decision-making. Showing the information about the reviewers, such as the number of reviews they have published \qian{(Q10, Mean = 4.75, SD = 1.89, 95\% CI [4.29, 5.18])} and the number of ``helpful'' votes those reviews received \qian{(Q11, Mean = 4.34, SD = 1.81, 95\% CI [3.90, 4.75])} are deemed relatively less useful than other information. Several participants reflected that they got used to filtering reviews by positive or negative aspects rather than using the information related to the reviewers. 

\qian{
Above all, we observed that participants are more aware of the potential bias after viewing data distribution behind user ratings with the help of the bias-aware design than those in the baseline condition. They were more aware of the difference of reviewers' experiences than other kinds of information based on the responses of questionnaire. In addition, participants who experienced our design gave positive feedback on its usability and usefulness. Most participants in the experimental group reported that the distribution of extreme emotions behind user ratings was the most useful information for them to make decisions.
}

\qian{\subsection{Strategies for Choosing Hotels (RQ2)}} \label{strategy}
To understand how people leverage our design to make decisions, we asked participants to share their strategies for selecting, eliminating and ranking hotels in the post-questionnaire. Together with the uploaded reasons in the systems (Fig. ~\ref{fig:baseline&Task} (D)), two authors coded the strategies using thematic analysis ~\cite{Virginia2006thematic} and finally reached consensus through two rounds of discussion. 
In general, we found that participants in both conditions share similar strategies of choosing hotels while they put different emphasis on several aspects.
We extracted five typical strategies from their feedback and summarized them in Table ~\ref{tab:reason}. We calculated the percentage of every strategy among all participants in each condition; four out of the five strategies occurred in both. We derive and report four salient findings regarding these strategies below.

\begin{table}[]
\caption{The coding scheme and summary of strategies from the participants in both conditions.}
\label{tab:reason}
\begin{tabular}{|c|c|c|}
\hline
{\color[HTML]{333333} \textbf{Strategies}}                               & \textbf{Condition} & {\color[HTML]{333333} \textbf{Percentages}} \\ \hline
                                                                & Prototype & 88.45\%          \\
\multirow{-2}{*}{Checking the Specific Aspects in User Reviews} & Baseline         & 94.23\%          \\ \hline
                                                                & Prototype & \textbf{55.77\%} \\
\multirow{-2}{*}{Considering Positive and Negative Reviews Collectively} & Baseline           & \textbf{28.85\%}                            \\ \hline
                                                                & Prototype & \textbf{13.46\%} \\
\multirow{-2}{*}{Referring to the User Rating Distribution}              & Baseline           & \textbf{48.08\%}                            \\ \hline
                                                                & Prototype & 17.31\%          \\
\multirow{-2}{*}{Referring to the Number of Negative Reviews}   & Baseline         & 13.46\%          \\ \hline
Referring to the Reviewers behind User Ratings                  & Prototype & \textbf{23.08\%} \\ \hline
\end{tabular}
\end{table}

\subsubsection{\textbf{Most participants in both groups selected hotels by checking the detailed aspects in user reviews.}} 
As shown in Table ~\ref{tab:reason}, 88.45\% and 94.23\% of participants in the two groups respectively mentioned they would like to inspect certain aspects of personal interests in the detailed user reviews when booking hotels. For example, P20 selected a hotel by checking several aspects she cared about.
\begin{displayquote}
    ``Cleanliness seems to be a consistent theme. I filtered reviews on cleanliness and looked for ones with the least amount of really negative reviews on this matter.'' -- P20 (F, 40, Prototype) 
\end{displayquote} 
P5 chose a hotel by checking the options for breakfast.
\begin{displayquote}
    ``Breakfast included vegetarian options. Breakfast is my favorite meal and I would not want to stay in a place where I would have to buy breakfast outside of the hotel.'' -- P5 (M, 25, Baseline)
\end{displayquote}
The difference is that participants who used our prototype system could filter the detailed aspects of reviews corresponded to the user rating bars while the participants in the baseline condition used tags to filter user reviews.

\subsubsection{\textbf{People tend to consider both positive and negative reviews together when using the prototype system.}}
The summary of strategies shows that participants who used our design were more likely to consider both positive and negative reviews \textbf{(55.77\%)} compared to those in the baseline group \textbf{(28.85\%)}. This provides evidence that our design can help users gain a comprehensive view of others' feedback on a hotel. For example, P5 chose Hotel 6 (J-shaped) and gave the following rationale: 
\begin{displayquote}
    ``I choose this hotel because the good to bad reviews ratio is optimal. The few bad reviews don't seem to bad to me.'' -- P5 (F, 20, Prototype)
\end{displayquote} 
P28 only used negative feedback to judge a hotel. 
\begin{displayquote}
    ``Since the prices were very similar, I strictly chose hotels based off the most negative reviews and the distribution of ratings.'' -- P28 (M, 33, Baseline)
\end{displayquote} 
However, in the fourth row of the Table ~\ref{tab:reason}, participants using the prototype referred to the number of negative reviews a little bit more than those using the baseline (17.31\% > 13.46\%). The reason is that participants were interested in the transparent information of the 1-point rating as they interacted more with the 1-point bar as shown in the Fig. ~\ref{fig:resultLog} (B).

\subsubsection{\textbf{People rely less on the rating distribution with the transparent design than those with the baseline.}} Although participants' selection of hotels in each group were affected by the rating distribution according to Fig. ~\ref{fig:resultLog}, in their feedback of strategies, we found that they have different degrees of dependence on the rating distributions. We could see that 48.08\% of participants filtered hotels by taking the user rating distribution into account in the baseline system, while the participants less mentioned the user rating distribution using our system (13.46\%) (Table ~\ref{tab:reason}). For people who chose hotels with the transparent design, some of them mentioned the rating distribution when they were aware of the potential bias. For instance, P62 (in the prototype group) mentioned that he eliminate the hotels with more 5-point ratings and 1-point ratings. 
\begin{displayquote}
    ``I eliminate hotels with U shape of comments or the ones have unacceptable negative comments.'' -- P62 (M, 28, Prototype)
\end{displayquote}
However, participants used the baseline relied on the rating distribution by considering either positive or negative reviews, such as P65 in the baseline group.
\begin{displayquote}
    ``If there's a high proportion of 5- and 4-star ratings, I am much more reassured that my experience will be good.'' -- P65 (F, 36, Baseline)
\end{displayquote}
The reason is that people can view the overall content of user reviews through the transparent information presented in our design. Thus, users are less likely to judge the quality of a hotel \textbf{solely} based on the rating distribution. 

\subsubsection{\textbf{The participants prefer to rely on the feedback of professional reviewers when choosing a hotel using the transparent design.}}
We discovered that 23.08\% of participants in the experimental condition (Table ~\ref{tab:reason}) used a new strategy not available in the baseline system; that is, taking the feedback from professional reviewers into consideration. 
P41 (in the prototype group) mentioned that 
\begin{displayquote}
    ``I chose these three hotels generally because they had the fewest one-star ratings given by the 'top' and 'pro' reviewers. I trust these peoples ratings more as they have experience giving reviews and consistently do so and are not likely to just give a bad rating based on one anomalous trip.'' -- P41 (M, 33, Prototype)
\end{displayquote}


\subsubsection{Summary of Operational Logs}
\qian{To explore how participants used our prototypes, we analyzed and compared the operational logs of all participants in both conditions. We plotted the result in Fig. ~\ref{fig:resultLog} (B) and (C).} 
As shown in Fig.~\ref{fig:resultLog} (B), the average number of clicks indicates how frequently users filtered reviews, while the average number of hovers implies how often they viewed relevant information by putting the cursor over it. 
Participants were more inclined to filter negative reviews in both systems and, compared to the baseline, participants who used our design hovered more times over the design bars as well as pies to view the information. We explored the possible reasons for this phenomenon from the responses in the post-study questionnaire, and found that participants who used our design preferred to get a general impression of a hotel by viewing the transparent information instead of reading a lot of user reviews. 
\begin{displayquote}
    ``I really want to use this design in the future as it will really help, firstly, not to waste your money, and secondly not to waste your time in searching for good hotels.'' -- P11 (F, 28, Prototype)
\end{displayquote}

The above findings are also confirmed by the average number of different kinds of interactions across the whole web site during the experiment (Fig.~\ref{fig:baseline&Task} (C)). More specifically, the total number of clicking and hovering interactions with our system far exceeded those of the baseline, while the number of scrolling up or down actions was considerably lower. This indicates that the participants with our design read fewer reviews than those in the baseline system, as scrolling the web page is necessary to see more user reviews that cannot be displayed within one screen. Note that the system interfaces are adaptive, so there is no need for users to scroll the page when looking at user ratings. 
Participants in the experimental condition also acknowledge in their post-questionnaire responses that they felt that our \qian{bias-aware} design was helpful for them to locate more relevant reviews conveniently.

\qian{To sum up, we found that participants in the experimental group used relatively different strategies for referring user ratings compared with the baseline condition. They tended to consider the information from different ratings more comprehensively and paid relatively less attention to any single type of feedback (e.g., negative feedback). Furthermore, participants in the experimental condition seemed to trust the feedback from professional reviewers more than other reviewers when they were informed about the distribution of reviewers' experiences behind user ratings.}

\qian{\subsection{Selection Result (RQ3)}}
\qian{We summarize the participants' final selections of hotels in Fig. ~\ref{fig:resultLog} (A). 
The bar chart shows the percentage of each hotel selected in both conditions respectively.} We grouped the hotels by the distribution type of user ratings (Fig. ~\ref{fig:Distribution}) and compared the participants' selections in the two conditions. 
We distilled several findings from participants' selection reasons by coding and analyzing the reasons provided by the participants using thematic analysis ~\cite{Virginia2006thematic}. 

\textbf{\begin{figure}
\centering
  \includegraphics[width=\columnwidth]{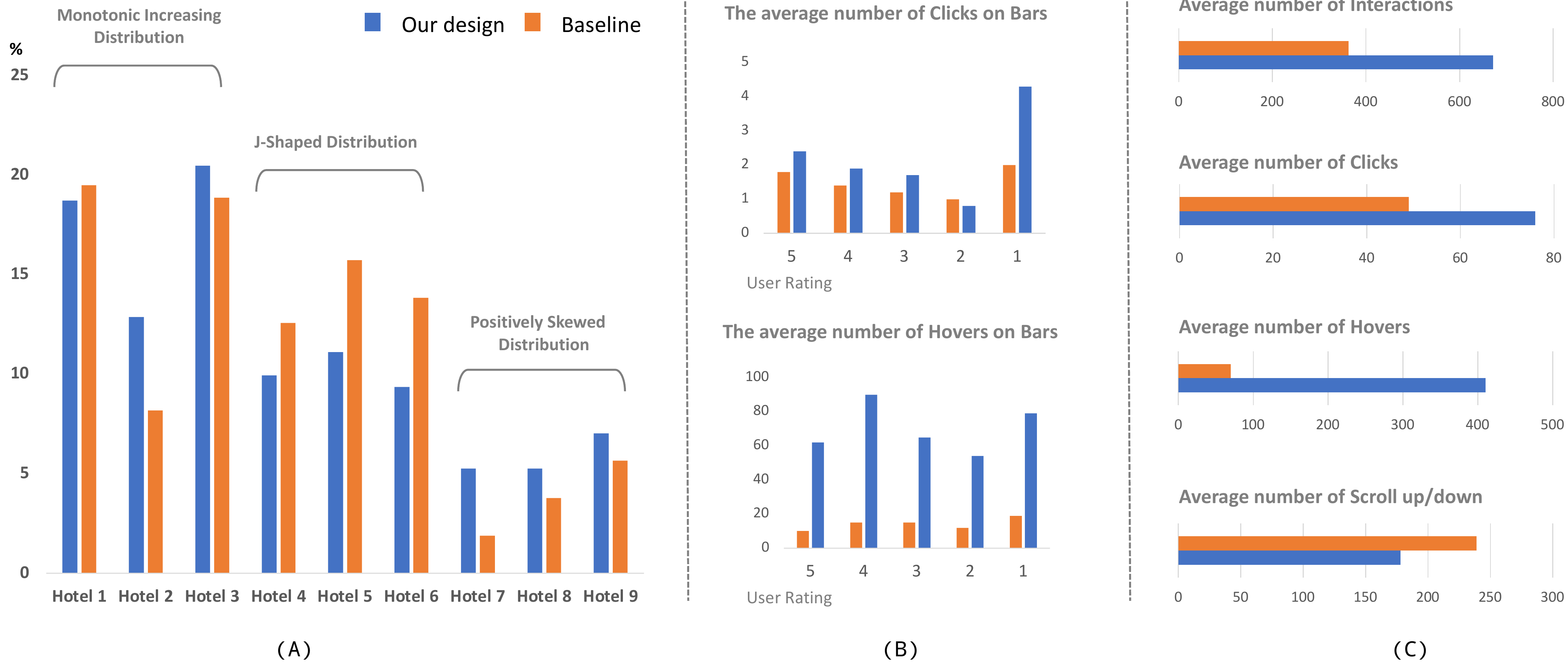}
  \caption{\qian{The statistical record of participants' choices of hotels and their interactions in the study. (A) shows the total number of times each hotel was selected by participants; (B) shows the average number of interactions on each rating bar for each participant in two systems. The x-axis represents the user ratings from 5 points to 1 point. The x-axis of (C) displays the average number of different types of interaction for each participant in two conditions.}
}  
  \label{fig:resultLog}
\end{figure}}

\subsubsection{\textbf{Participants' choices of hotels were affected by the distribution of user ratings.}}
\qian{As shown in Fig. ~\ref{fig:resultLog} (A),} the hotels with a monotonic increasing distribution of user ratings got the most votes, while hotels with a positive skewed distribution of user ratings were the least favorable to our participants. The hotels with a J-shaped user rating distribution gained moderate votes. 
Therefore, people's decisions could be affected by the distribution of user ratings as the participants tended to choose hotels with more 5-point and fewer 1-point ratings. 

\subsubsection{\textbf{Seeing the transparent information behind the user ratings could affect the selection of hotels.}}
The other finding is that, in Fig. ~\ref{fig:resultLog} (A), \qian{less percentage of participants using the bias-aware design} chose the No.4 - No.6 hotels under the J-shaped user rating distribution (the typical biased distribution caused by the self-selection bias) than those using the baseline. Moreover, more participants chose the hotels with a positively skewed rating distribution (the distribution with the least bias \cite{lim:17mitigating}) using the prototype system than those with the baseline.
We analyzed the reported reasons for selection elicited by the participants. We discovered that, compared to people applying the baseline system, participants in the experimental condition considered the general rating distribution more when they could see the transparent and more detailed information behind different ratings (1–5 points). 
To be more specific, they would like to inspect and analyze the auxiliary data shown in the pie charts under the user rating bars (Fig. ~\ref{fig:prototype} (B)) and make decisions based on it. 
\begin{quote}
    ``I selected hotels according [to] the most balanced of all aspects of ratings to reviewers and contributors in the pies.'' -- P12 (M, 30)
\end{quote}

\qian{
Generally, compared with the baseline, a higher percentage of participants in the experimental group selected hotels with a less biased distribution of user ratings (i.e., positively skewed distribution \cite{lim:17mitigating}), and a smaller percentage of them chose the hotels with the potentially biased distribution (e.g. J-shaped distribution).
}


\section{Discussion}
The user study results indicate the effectiveness of our design for raising awareness of \minor{the self-selection} bias and facilitating people's decision-making process. 
In this section, we discuss lessons learned from the implementation of design and user study. We also reflect on the implications for supporting people's decision-making by raising awareness of the self-selection bias through the an informed and unintrusive way.

\subsection{Effect of Information Transparency}
We decided the exact transparent information by a formative study that collected the critical information people care about and people's perceptions of the transparent information. In the study, we showed the information combining user ratings with reviews to people and explore how they exploit the information and make decisions. The results indicate that the information provided are helpful for decision-making in general (Fig. ~\ref{fig:questionnaireRate}) while changing people's behaviors and their strategies of referring user ratings and reviews for decision-making (Section ~\ref{strategy}). 

However, it is necessary to mention that people in the group with the \qian{bias-aware} design spend 10 more minutes on average than the baseline in the study. The reason was that, on the one hand, the experimental group needed to watch 1 more minute of the introduction video to understand our design and task before the study (Section ~\ref{studyprocedure}), and they answered more questions in the post-questionnaire. On the other hand, the operation logs shows that they interacted more with the \qian{bias-aware} design than the baseline (Fig. ~\ref{fig:resultLog}). Hence, how much and the granularity of transparent information to be shown are all needed to be considered. If we disclose too much information, users may abandon checking the details as they can be overwhelmed. For example, while other aspects of user ratings and reviews may be biased (e.g., reviewers' gender and ethnic group), the formative study results suggest that they are not the key concerns of most crowd respondents during hotel selection. 
Rather than presenting all potentially biased dimensions of the user-generated data, we only emphasize on the most salient ones that have strong impact on people's decisions.

In addition, the process of making information transparent is not just about the information itself. It also concerns how to present the information that was previously ``invisible'' to users without overloading users. It is thus necessary to consider and reflect the information needs of users in the design.
We investigated how to present transparent information intuitively based on the formative study and the discussion with two visualization experts. Then, we proposed three alternatives and compared how people perceive and explain the information in these designs. While some of the alternatives could better support visual analytic tasks, they are more visually complex, more challenging to interact with, more space-consuming, and harder to scale. Based on user feedback in the pilot study, we selected the one design that is the most acceptable to our target users (i.e., lay public) and improved it by adding user-friendly interactions. 
This process ensured that at least most ordinary users could easily learn to use our design and grasp the conveyed information quickly and freely in their decision-making process.
In general, designs related to information transparency should always consider the needs and characteristics of users and strike a balance between information transparency and information complexity.

\subsection{Evaluation of Awareness}
Certain biases in user-contributed contents (e.g., imbalance in gender) could be alleviated by applying a better sampling technique proactively, such as the methods used by previous works \cite{Kemal:20Manipulation, lim:17mitigating, wu:17mitigating, nagtegaal:20designing, Askalidis:17overcoming}, as discussed in the Related Work (Section ~\ref{lr2.2}).  
However, some other types of biases in user-generated data are caused by unconsciousness, such as people's extreme emotions ~\cite{schoenmuller:19extreme}, their knowledge/expertise ~\cite{halevy:19ubiquity}, and selection of individuals ~\cite{Bhole:17Self-SBias}. They tend to be more implicit and may not be easily detected, measured, and mitigated by algorithms. 
Therefore, in this work, we chose one kind of implicit bias (i.e., self-selection bias) and proposed to raise people's awareness of the bias as the goal of our work ~\cite{baeza:18bias}, on the assumption that such biases could not be technically eliminated from the data. 

One challenge we met in this work is the evaluation of the people's awareness, as it is not proper to directly ask whether they are aware of the bias or not. A previous study by psychologists has shown that people tend to appear knowledgeable when their awareness is explicitly tested ~\cite{srivastava:16effects}. We explored the literature and found that the way of evaluating the awareness should be associated with the concrete task and goal. For example, a recent work by CSCW researchers used the reliability scores in the pre- and post-survey to measure the awareness of the echo chamber effect ~\cite{jeon2021chamberbreaker}. Another work by Eslami et al. ~\cite{eslami2017careful} tested users' awareness of the bias in algorithms by labeling whether they could articulated a discrepancy between their intended review score and the system output. 
We designed the questions in the post-questionnaire based on the implicit association test (IAT) in psychology ~\cite{nosek2005understanding} and the questions are adapted with the defined transparent information related to the self-selection bias. 
Therefore, the method of evaluation of people's awareness should be carefully considered and designed, especially for the awareness of the implicit bias.

Additionally, there is an interesting finding of the awareness testing result in Fig. ~\ref{fig:questionnaireRate}. People were more aware of the potential biased reviewers' experience than the extreme emotions and various aspects in user reviews while the emotion is the most care than the other two types of information according to the formative study (Section ~\ref{formative study}). We attribute this to people's belief updating problems as people have a demand of the extreme user reviews (e.g., negative reviews) to some extent for making decisions ~\cite{ambuehl2018belief}. 
Hence, although people may be aware of the bias in the extreme feedback, they still tend to associate the user rating score (e.g., 1-point) with the corresponding emotions (negative only). Moreover, we also attribute this to one limitation of our work in the evaluation of people's awareness. The questions for evaluating awareness in the post-questionnaire could be improved by gathering people's reactions and perceptions of the aware-testing questions in the pre-study session or pilot study. 
Therefore, evaluating people's awareness of bias needs to consider implicit question design, specific application scenarios and pre-estimate people's responses to the evaluation method.

\subsection{Unintrusive Design for Awareness-raising and Informed Decisions}

In our design process, we decided to use a unintrusive way to raise people's awareness of the self-selection bias. 
The reason is that the direct warning of potential bias that explicitly informs people can be a kind of inducement for users which may cause unwanted emphasis on the data rather than the decision ~\cite{Law2021CausalQA}. 
Therefore, we did not add any explicit prompts or warnings of the self-selection bias in the design, and we even did not reveal the goal of our study or design during the experiment. 
The study results show that this implicit approach can effectively increase people's knowledge and perception of possible biases in user ratings and reviews without distracting them from their decision process during the experiment. 
However, in the study results (Fig. ~\ref{fig:questionnaireRate}), at least for the awareness of the different aspects (I3), it seems that people's general level of awareness has not reached the ideal level. One reason may be the implicit way of the evaluation and the design. Hence, it deserves further exploration of the effects of implicit design, as well as how to increase awareness in the implicit way. 

In addition, it remains to be studied on how to efficiently offering the awareness-raising design for informed decisions in daily life. The latest research in the visualization domain proposed a method to help people realize their biased behavior during the data analysis process by showing the interaction logs explicitly in visual analytic systems ~\cite{narechania2021lumos, wall2021left}. We could adopt this way in the online decision-making scenario by making people's behaviors transparent (e.g., scanning extreme reviews) and informing them about the potentially biased behavior on the fly. 
From the perspective of applying visualization, the literacy of visualization is key point that should be considered by designers as ordinary users may need some efforts to understand and interpret the information of the design.


\subsection{Limitations and Future Work} 
We acknowledge that this work has several limitations. 
One limitation is that the sentiment analysis results and the keywords from user reviews involved in our experiment were all processed by automatic algorithms. Even though we chose a widely used tool ~\cite{Gardner2017AllenNLP} and a powerful model ~\cite{grootendorst2020keybert}, we still cannot guarantee the algorithms are unbiased. 
These are emergent tools for checking model fairness ~\cite{bellamy2019ai}, and we hope to ensure the use of unbiased algorithms with these tools before integrating them into the backend of our system in the future. 

Secondly, for the purpose of our controlled study, variables other than reviews and derived information are confined to a narrow range (e.g., price) to limit potential confounding effects. 
In a real world situation, users may face a wide variety of available hotels and their choice may be largely influenced by other factors, such as price, location, and even images of the hotel rooms. 
Moreover, considering the duration of the experiment and the burden on users, we only provided nine alternative hotels, while people may browse much more candidates in reality. 
We consider assessing people's awareness of the self-selection bias in more complex situations, while using more hotels to evaluate the design in future work.

Thirdly, as we conducted our study using an online crowdsourcing platform, there are some uncertainties in this setup, such as unstable internet connections, browser incompatibilities, and other technical problems. To reduce the impact of this, we implemented some detailed functions to optimize the user experience, such as the adaptive interface design, prompt windows in the system, etc. While these helped in the experiment, several participants said they experienced compatibility and network issues. Future work needs to provide a more robust system for evaluating the design with a broader audience. 
Additionally, a long-term study is also necessary for observing how users leverage the \qian{bias-aware} design for making decisions based on online ratings/reviews in real situations.

Despite these limitations, the user study does help us to learn more about users and their experiences, identify possible shortcomings in our design, and pinpoint improvements for our future work. 
\qian{In addition, exploring the impact of COVID-19 on the user-generated content and the possible bias behind (e.g., ratings and reviews of products or services) could be an interesting future direction to explore. We will also explore if our proposed approach can be generalized to other real-life scenarios in which user-generated content may affect users’ decisions (e.g., buying a product online) in the future.} As such, we will continuously improve our work that aims to raise people's awareness of various biases, and reflexively apply the design in other scenarios (e.g., buying products) based on the broader HCI and CSCW community’s feedback.

\section{Conclusion}
We proposed a \qian{bias-aware} design for user ratings with the aim of raising people's awareness of the self-selection bias. 
The design shows the proportions of three kinds of information that are related to the self-selection bias while affecting people's decision-making when referring user ratings/reviews. To evaluate whether the design could increase people's awareness and help them make decisions, we conducted an online study through a crowd-sourcing platform with 136 participants. The results show that the \qian{bias-aware} design can significantly increase people's awareness compared to the baseline, and people can use the design more efficiently through interaction, thus facilitating their decision-making process.
We analyzed several key points obtained from this work, including transparency information, evaluation of people's awareness of bias, and the non-intrusive design. Future work may further explore these points under various tasks or scenarios of decision-making. We hope this work can inform and inspire designers and researchers in the broad HCI communities to investigate bias from the end user's perspective.

\section{Acknowledgement}
This work is partially supported by the Research Grants Council of the Hong Kong Special Administrative Region, China under General Research Fund (GRF) with Grant No. 16203421.

\bibliographystyle{ACM-Reference-Format}
\bibliography{sample-base}

\received{January 2022}
\received[revised]{April 2022}
\received[accepted]{August 2022}

\end{document}